\documentclass{article}

\usepackage[english]{babel}

\usepackage[letterpaper,top=2cm,bottom=2cm,left=3cm,right=3cm,marginparwidth=1.75cm]{geometry}

\usepackage{amsmath}
\usepackage{graphicx}
\usepackage{multirow}
\usepackage{hhline}
\usepackage{amsmath,amsfonts}
\usepackage{algorithmic}
\usepackage{algorithm}
\usepackage{array}
\usepackage[caption=false,font=normalsize,labelfont=sf,textfont=sf]{subfig}
\usepackage{textcomp}
\usepackage{stfloats}
\usepackage{url}
\usepackage{verbatim}
\usepackage{graphicx}
\usepackage{rotating}
\usepackage{tikz}
\usetikzlibrary{positioning}
\usetikzlibrary{patterns}
\usetikzlibrary {shapes.symbols}
\usetikzlibrary {shapes.geometric}
\usepackage{multirow}
\usepackage{makecell}
\usepackage{threeparttable}
\usepackage[numbers]{natbib}
\usepackage{pgf}
\usepackage{amsmath}
\usepackage{amsthm}
\usepackage{amssymb}

\newtheorem{theorem}{Theorem}        
\newtheorem{lemma}[theorem]{Lemma}   
\newtheorem{definition}{Definition} 

\usepackage[colorlinks=true, allcolors=blue]{hyperref}

\title{A Model-free Biomimetics Algorithm for Deterministic Partially Observable Markov Decision Process}

\author{
  Yide Yu\thanks{Macao Polytechnic University, Macao, China} \\
  \texttt{yide.yu@mpu.edu.mo} \\
  \and
  Yue LIU\footnotemark[1] \\
  \texttt{yue.liu@mpu.edu.mo} \\
  \and
  Xiaochen YUAN\footnotemark[1] \\
  \texttt{xcyuan@mpu.edu.mo} \\
  \and
  Dennis WONG\footnotemark[1] \\
  \texttt{cwong@mpu.edu.mo} \\
  \and
  Huijie LI\footnotemark[1] \\
  \texttt{huijie.li@mpu.edu.mo} \\
  \and
  Yan MA\thanks{Beijing University of Posts and Telecommunications, Beijing, China} \\
  \texttt{mayan@bupt.edu.cn}
}

\begin{document}
\maketitle

\begin{abstract}
Partially Observable Markov Decision Process (POMDP) is a mathematical framework for modeling decision-making under uncertainty, where the agent's observations are incomplete and the underlying system dynamics are probabilistic. Solving the POMDP problem within the model-free paradigm is challenging for agents due to the inherent difficulty in accurately identifying and distinguishing between states and observations. We define such a difficult problem as a DETerministic Partially Observable Markov Decision Process (DET-POMDP) problem, which is a specific setting of POMDP. In DET-POMDP, states and observations are many-to-one relationships. In this setting, the state is obscured, and its relationship is less apparent to the agent. Moreover, this setting creates more obstacles for the agent to infer the state through observations. To effectively address this problem, we convert DET-POMDP into a fully observable MDP using a model-free biomimetics algorithm BIOMAP. BIOMAP is based on the MDP-Graph-Automaton framework to distinguish authentic environmental information from fraudulent data. Thus, it enhances the agent’s ability to develop stable policies against DET-POMDP. The experimental results highlight the superior capabilities of BIOMAP in maintaining operational effectiveness and environmental reparability in the presence of environmental deceptions when comparing with existing POMDP solvers. This research opens up new avenues for the deployment of reliable POMDP-based systems in fields that are particularly susceptible to DET-POMDP problems.
\end{abstract}

\section{Introduction}

\noindent The DETerministic Observable Markov Decision Process (DET-POMDP)~\cite{bonet2012deterministic} problem is an emerging area within the domain of Partially Observable Markov Decision Processes (POMDPs). Despite having a more limited problem coverage domain than the traditional POMDP due to its deterministic state transition and observation functions, the DET-POMDP is still applicable to many real-world scenarios and remains difficult to solve under the model-free setting. The challenge arises when states are concealed under observations, making it difficult for the agent to recognize changes in the environment when it solely relies on these observations. This issue delves into the implications of decision-making accuracy, exploring the challenges and potential consequences for both researchers and practitioners. For instance, medical diagnosis problem \cite{hauskrecht1997planning} highlights the possibility of the medical equipment providing inaccurate readings to doctors. Given that no equipment can fully capture the actual state of patients, even if two patients exhibit the same readings on the equipment, they may have individual differences. By considering the medical equipment as the environment in a POMDP, it becomes evident that the environment may mislead doctors in their medical decision-making, leading to a DET-POMDP problem.

It is crucial to note that in reality, agents are unable to explicitly model the environment in advance. This limitation arises due to the high dimensionality of the state space, making it challenging to accurately capture it based solely on human perception. Despite having access to a dataset reflecting the environment, it is impossible to record every aspect of the real environment. This dilemma leads to the \textit{Cognitive Fog} phenomenon which misleads the decision relies on observations. Consequently, the constructed environment model based on the dataset remains partially observable, leading to inaccuracies in the modeling process. As a result, addressing the POMDP problem within the paradigm of model-free approaches becomes particularly challenging.

Based on our survey, the majority of existing POMDP solvers are predominantly model-based. These algorithms are more inclined towards establishing the relationship between states and observations when an environment model is available. However, in real-world scenarios, it is common to encounter situations where no environment model is present priorly. As a result, the applicability of model-based algorithms becomes limited in practice. Furthermore, many POMDP solvers rely on the belief space technique \cite{kaelbling1998planning, platt2010belief}. Constructing a comprehensive belief space is difficult when there is a lack of information about states and observations \cite{cassandra1998survey}, especially when employing model-free methods \cite{silver2010monte}. Indeed, solving POMDP problems using model-free approaches is a realistic and pressing challenge that needs to be addressed. It is worth noting that most model-free algorithms fall into the category of reinforcement learning \cite{mnih2015human, mnih2013playing}, but most of the reinforcement learning algorithms are assumed to be effective under fully observable Markov Decision Processes (MDPs), ignoring partially observable problems. Therefore, designing a model-free algorithm to solve the POMDP problem is a tricky problem.

\begin{figure}[ht]
  \centering
  \resizebox{0.9\textwidth}{!}{%
    \begin{tikzpicture}[node distance=2cm, auto]
      \node[draw, rectangle, fill=red!70] (obs1) {\textcolor{black}{$o_1$}};
      \node[draw, rectangle, below=1cm and 0cm of obs1, fill=red!30] (obs2) {\textcolor{black}{$o_2$}};
      
      \node[draw, circle, above right=0.5cm and 1.5cm of obs1, node distance=3cm, fill=red!10] (state1) {\textcolor{black}{$s_1$}};
      \node[draw, circle, below of=state1, node distance=1cm, fill=red!10] (state2) {\textcolor{black}{$s_2$}};
      \node[draw, circle, below of=state2, node distance=1cm, fill=red!10] (state3) {\textcolor{black}{$s_3$}};
      \node[draw, circle, below of=state3, node distance=1cm, fill=red!10] (state4) {\textcolor{black}{$s_4$}};

      \draw[->] (state1) -- (obs1) node[midway, left] {\tiny $90\%$};
      \draw[->, dashed] (state1) -- (obs2) node[pos=0.1, right] {\tiny $10\%$};
      \draw[->] (state2) -- (obs1) node[pos=0.4, right] {\tiny $65\%$};
      \draw[->, dashed] (state2) -- (obs2) node[pos=0.2, right] {\tiny $35\%$};
      \draw[->] (state3) -- (obs1) node[pos=0.001, left] {\tiny $15\%$};
      \draw[->, dashed] (state3) -- (obs2) node[pos=0.3, right] {\tiny $85\%$};
      \draw[->] (state4) -- (obs1) node[pos=0.1, left] {\tiny $50\%$};
      \draw[->, dashed] (state4) -- (obs2) node[pos=0.2, left] {\tiny $50\%$};
      
      \node[draw, rectangle, right of= obs1, node distance=3cm, fill=red!50] (obs3) {\textcolor{black}{$o_1$}};
      \node[draw, rectangle, below=1cm and 0cm of obs3, fill=red!50] (obs4) {\textcolor{black}{$o_2$}};
      \node[draw, circle, above right=0.5cm and 1.5cm of obs3, node distance=3cm, fill=red!10] (state5) {\textcolor{black}{$s_1$}};
      \node[draw, circle, below of=state5, node distance=1cm, fill=red!10] (state6) {\textcolor{black}{$s_2$}};
      \node[draw, circle, below of=state6, node distance=1cm, fill=red!10] (state7) {\textcolor{black}{$s_3$}};
      \node[draw, circle, below of=state7, node distance=1cm, fill=red!10] (state8) {\textcolor{black}{$s_4$}};

      \draw[->] (state5) -- (obs3) node[midway, left] {\tiny$100\%$};
      \draw[->] (state6) -- (obs3) node[pos=0.3, left] {\tiny$100\%$};
      \draw[->, dashed] (state7) -- (obs4) node[pos=0.3, left] {\tiny$100\%$};
      \draw[->, dashed] (state8) -- (obs4) node[pos=0.3, left] {\tiny$100\%$};


      \draw[dashed] ([shift={(-0.05cm,-2.4cm)}]obs1.west) rectangle ([shift={(0.8cm,0.5cm)}]state1.north west);  
      
      \draw[dashed] ([shift={(-0.05cm,-2.4cm)}]obs3.west) rectangle ([shift={(0.3cm,0.5cm)}]state5.north east);

      \node[above right, xshift=-58pt, yshift=1.5pt] at (state1.north west) {\tiny \textbf{General POMDPs}};
      \node[above right, xshift=-45pt, yshift=1.5pt] at (state5.north west) {\tiny \textbf{DET-POMDP}};
      
    \end{tikzpicture}%
  }
  \caption{General POMDP vs. DET-POMDP example. In DET-POMDP with environment-free modeling, the correspondence between states and observations is more challenging compared to general POMDPs. This difficulty arises from the unchanging pattern of occurrence of observations, which hampers the process of establishing a reliable mapping between states and observations.}
  \label{fig: environmental fraud}
\end{figure}
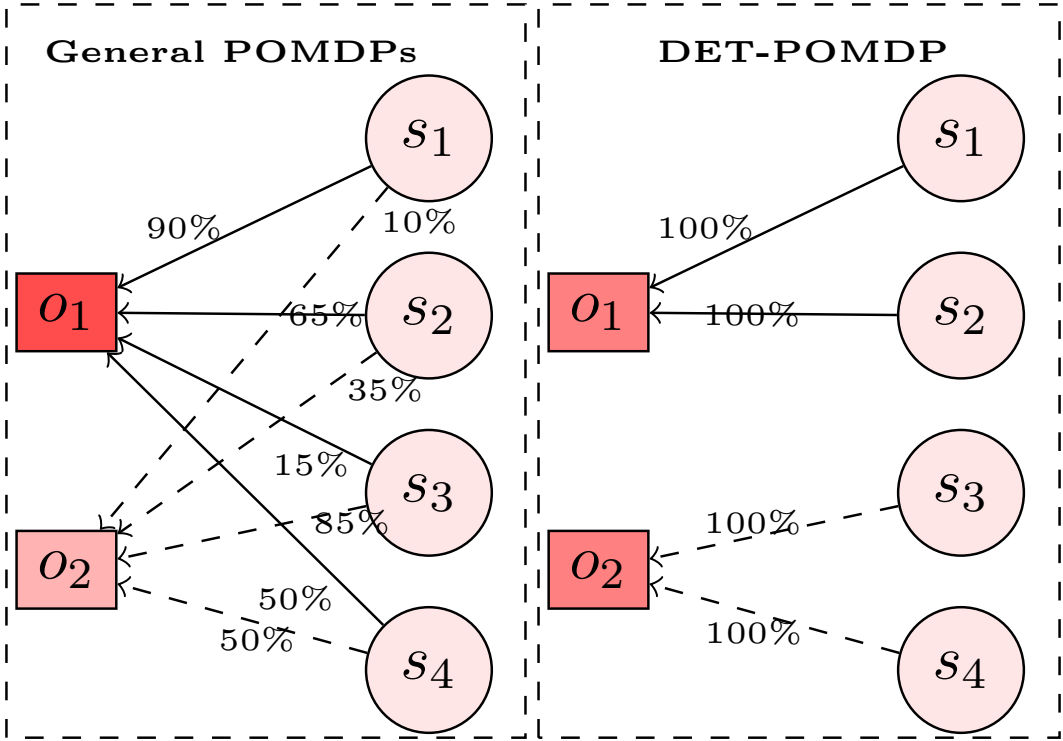

To solve the POMDP problem based on the model-free paradigm, first we introduce the DET-POMDP problem. Fig. \ref{fig: environmental fraud} illustrates that in our study, DET-POMDP handles the scenario where $s_1$ and $s_2$ solely correspond to $o_1$. This contrasts with the general POMDPs framework, where the likelihood of observation is less than $100\%$. When the agent is unaware of the specific states, encountering the same observations may lead the agent to assume that they are identical. However, agents are often in this misdirection and don't realize it. Even if the agent knows the predetermined number of states and observations, it still faces challenges in distinguishing the correspondence between states and observations due to the fixed state transfer probability. After that, we propose a nature-inspired biomimetics algorithm named BIOMAP, which draws inspiration from the path integration ability observed in desert ants. Desert ants possess the ability to remember the route from their nest to food sources. This inspires us to rely not solely on environmental information but also on trusted actions of the agent. We construct a MDP-Graph-Automaton framework as the theoretical foundation of BIOMAP. BIOMAP creates a \textit{Compact Action Graph} to retain the action trajectory and employs a \textit{Markov Automaton} to determine whether the environment is fully or partially observable. Finally, after converting DET-POMDP to MDP, the optimal policy is then obtained on the graph using the shortest path algorithm.

In our research, we have made three notable contributions:

\begin{enumerate}
    \item We conducted a quantitative analysis of the Cognitive Fog phenomenon in DET-POMDPs, introducing Q-value variance as a measure of decision-making bias caused by states overlap. This theoretical contribution provides insights into the challenges of policy optimization under DET-POMDP.
    \item We designed the MDP-Graph-Automaton framework, which exhibited a complete triple mapping from MDP to graph to Automaton. We provided a theoretical foundation for this framework, and the framework presented a new representation of POMDP solvers.
    \item Our research resulted in the development of the BIOMAP algorithm. After the results, it was verified that the BIOMAP algorithm, as model-free, was as accurate as some model-based algorithms in solving the DET-POMDP scenario accurately. This algorithm represented a groundbreaking solution for addressing the DET-POMDP problem based on a model-free approach and served as a catalyst for further advancements in our field.
\end{enumerate}

This paper is divided into three main sections. In Sec. \ref{sec: related works}, we investigate existing POMDP solvers for traditional POMDPs. Moving on to Sec. \ref{sec: methodology}, we introduce the MDP-Graph-Automaton framework, which establishes the connection between the trilateral relationship. We propose the BIOMAP algorithm and analyze its time complexity. Additionally, we discuss the bionic process of desert ants. In Sec. \ref{sec: experiment} presents the design of a simulator called \textit{Mask Cliff Walking}, which allows us to validate the performance of BIOMAP and compare it with other traditional POMDP solvers. Moreover, we present the results of our experiments, demonstrating that BIOMAP outperforms most of the other model-based solvers on the Mask Cliff Walking problem.

\section{Related Works} \label{sec: related works}

As a complete POMDP ecosystem, POMDPs.jl \cite{egorov2017pomdps} serves as the foundation for our investigation. We focus our research within the scope of POMDPs.jl, leveraging its comprehensive functionalities and resources to explore and analyze various aspects of POMDPs. By utilizing this powerful framework, we aim to gain insights and develop innovative solutions within the realm of POMDPs.

The related works can be categorized into four distinct categories. In Tab. \ref{tab: related works}, we summarize the common advantages and disadvantages of each category. However, it is important to note that none of these categories provide an efficient solution for addressing the problem of DET-POMDP. Consequently, we are unable to discuss DET-POMDP within the context of any of these existing categories. In addition to inconsistent environmental complexity \cite{10495179}, observability plays a significant role in distinguishing between difficult and simple decision problems.

\subsection{Value episode-based Algorithms}

Algorithms based on Value episodes utilize value episodes to update belief states and use this information to determine the optimal policy in POMDPs. However, these algorithms suffer from the Curse of Dimensionality when applied to real-world problems with large state or action spaces. Specifically, the presence of a large observation space can result in a multitude of possible observation outcomes. Consequently, the computational cost increases significantly until convergence is achieved due to the direct correlation between large state space and computational requirements.

The QMDP algorithm \cite{littman1995learning} is based on the assumption that any uncertainty in the agent's current belief state will be eliminated after the next action. However, this assumption disregards the uncertainty associated with future states. Despite this limitation, QMDP utilizes the value function of an MDP at each step. Belief Grid Value episode \cite{lovejoy1991computationally} approximates the continuous belief space by decentralizing a set of belief points. It then updates the value of these belief points using the value episode algorithm, similar to QMDP. Similarly, Point-based value episode (PBVI) \cite{pineau2003point, shani2013survey} selects a finite set of representative belief points and updates their value function. Incremental Pruning \cite{cassandra2013incremental} reduces computational complexity by employing dynamic programming and pruning policies that have minimal contributions to the value function.

\subsection{Monte Carlo Algorithms}

Monte Carlo algorithms are designed to leverage the power of Monte Carlo estimation through random sampling and simulation of future paths. However, these algorithms are not without their limitations. One prominent disadvantage is the issue of sample dependence, which can introduce bias if the samples fail to adequately represent the entire space. Consequently, Monte Carlo algorithms require a large number of samples to mitigate this drawback. Additionally, these algorithms face significant computational requirements, with the sample space growing exponentially as the dimensions increase. Consequently, when applied to high-dimensional problems, Monte Carlo algorithms can be computationally expensive and time-consuming. Another common shortcoming is the presence of variance issues. Since Monte Carlo methods rely on random sampling, insufficient samples can lead to substantial and unstable variance, further impacting the reliability of the results.

The Partially Observable Upper Confidence bound applied to Trees (PO-UCT) algorithm \cite{silver2010monte, kocsis2006bandit} combines two techniques: Upper Confidence bounds for Trees (UCT) \cite{kocsis2006bandit} and Monte Carlo Tree Search (MCTS) \cite{chaslot2008monte}. PO-UCT constructs a search tree on the belief space, where nodes represent belief states and edges represent actions. The child node corresponds to the successor belief state of the parent node. PO-UCT evaluates the expected performance of each action and gradually builds an approximately optimal tree. Partially Observable Monte Carlo planning with observation widening (POMCPOW) \cite{sunberg2018online} is an extension of MCTS. POMCPOW incorporates two essential mechanisms for solving POMDPs. The first mechanism is observation widening, which groups observations into classes instead of creating nodes for each observation. This reduces the tree width and enables efficient search. The second mechanism is the use of belief nodes. POMCPOW maintains belief states for each node, representing the probability distribution over environmental states. This design allows POMCPOW to better evaluate the outcomes of actions using prior information. In contrast, Monte Carlo Value episode (MCVI) \cite{bai2014integrated} randomly samples states and updates them in the state space, rather than updating and evaluating the value function across the entire state space. Particle Filter Trees with Double Progressive Widening (PFT-DPW) \cite{sunberg2018online} is a hybrid algorithm that combines the features of Particle Filter and MCTS. The DPW strategy gradually increases the number of sub-nodes to avoid node explosion in large or continuous action spaces. This strategy adds a new action to explore when the previous action has been explored sufficiently. PFT-DPW is a powerful tool for solving complex problems with a large number of possible actions.

\subsection{Tree Search-based Algorithms}

The use of tree structures to solve Markov model families is a hot direction, such as sharing information in Markov chains on tree structures \cite{10399895}. Tree Search-based algorithms are commonly employed to find optimal solutions by constructing a decision tree. In this tree, each node represents a state, while branches symbolize actions. These algorithms navigate the tree to identify a sequence of actions that lead to the best outcome. However, managing the branching factor poses a challenge, as a large number of branches require significant computational resources and time. Additionally, when dealing with immense search spaces, Tree Search-based algorithms cannot guarantee optimality.

The Fast Informed Bound (FIB) algorithm, described in \cite{kochenderfer2022algorithms}, efficiently approximates an upper bound for the optimal value function. This approximation aids in evaluating the maximum return, serving as a crucial component in guided search. By accelerating the search process, it facilitates the rapid discovery of solutions. Successive Approximations of the Reachable Space under Optimal Policies (SARSOP) \cite{kurniawati2009sarsop} extends the reachable belief space to search for the optimal policy within it. It considers that certain belief spaces are not unreachable under the optimal policy. Anytime Regularized DEterminized Sparse Partially Observable Tree (AR-DESPOT) \cite{somani2013despot} focuses on tackling significant POMDP problems with large state or action spaces. It constructs a DESPOT decision tree that is sparse, deterministic, and regularized. The algorithm utilizes this tree to approximate the belief space and the optimal policy. Adaptive Online Packing-guided Search (AdaOPS) \cite{wu2021adaptive} employs two methods for solving POMDPs. Firstly, it utilizes an adaptive particle filter technique, which employs a set of particles to represent the state space of the environment. This technique allows for flexible adjustment of the number and distribution of particles. Secondly, AdaOPS incorporates the fusion of observation branches to strike a balance between variance and bias in estimations.

\subsection{Heuristic and Approximation Algorithm}

Heuristic and approximation algorithms leverage previous experiences or principles to expedite problem-solving by reducing the search space. However, these algorithms cannot guarantee global optimality due to their reliance on approximation and heuristic methods. Furthermore, heuristic methods heavily depend on domain knowledge and the design of the heuristic function.

Anytime Error Minimization Search (AEMS) \cite{ross2007aems} is a technique that utilizes offline knowledge to guide online searching. At each step, the agent makes decisions based on the current belief state, and AEMS explores the space to identify the optimal policy. As new observations are acquired, AEMS updates its beliefs and incorporates them into the search process. This approach aims to minimize errors and enhance the efficiency of the search.

\section{Methodology} \label{sec: methodology}

In this section, our focus is on providing a solution using a model-free algorithm for POMDPs. To begin, in Sec. \ref{sec: problem statement}, we present the DET-POMDP problem and the harm of DET-POMDP. Moving on to Sec. \ref{sec: Biotechnology}, we describe the bionic abstraction process of mapping desert ants to BIOMAP. In Sec. \ref{sec: framework}, we construct the structure of MDP-Graph-Automaton. However, we acknowledge that the MDP-Graph-Automaton framework still encounters challenges in determining boundaries accurately. To address this issue, we propose the Boundary Arbiter pseudo-algorithm in Sec. \ref{sec: boundary}, which aims to solve the problem of boundary definition effectively. Furthermore, in Sec. \ref{sec: BIOMAP}, we provide the pseudo-algorithms for BIOMAP, alongside an analysis of its time complexities. Understanding the computational requirements of these algorithms is essential for evaluating their practical feasibility in Sec. \ref{sec: time complexity}.

\subsection{Problem Statement and Objective} \label{sec: problem statement}

Two standard mathematical models for Reinforcement Learning are MDP and POMDP, which are used to model and solve decision problems. MDP, as described by Puterman \cite{puterman2014markov}, is a mathematical framework that effectively captures the decision-making process and characterizes the state transition process.

However, in the real world, agents often cannot achieve the true state, especially when the cardinality of the observation space differs from that of the state space. In such cases, agents receive observations that do not directly correspond to the actual states. This scenario is characterized by POMDPs~\cite{kaelbling1998planning}, which extend standard MDPs. In POMDPs, agents lack complete visibility of the environment's states and can only observe partial information. These uncertain observations create ambiguity in determining the optimal decision.

This research focuses on the DET-POMDP model~\cite{bonet2012deterministic}, which is a sub-class of POMDPs. \textbf{First, we introduce the DET-POMDP model.}

\subsubsection{DET-POMDP}



DET-POMDP refers to the observation function and state transition function being deterministic in the POMDP. We modify the definition of DET-POMDP in the symbolic system of the POMDP in Def. \ref{def: env}.

\begin{definition}[DET-POMDP model~\cite{bonet2012deterministic}]
    A DETerministic Partially Observable Markov Decision Process (DET-POMDP) is defined as a tuple consisting of: a finite state space $\mathcal{S}$, finite sets of applicable actions $\mathcal{A}_s$ for each state $s$, a finite set of observations $\mathcal{O}$, an initial state subset or distribution $\mathcal{S}_0$, a subset of goal states $\mathcal{S}_g$, a \textbf{deterministic} transition function $T(s, a)$ that specifies the resulting state after applying an action, a \textbf{deterministic} observation function $\Omega(s, a)$ that defines the observation received upon entering a state after an action, and the reward function $R(s, a)$ representing the immediate reward of an action in a given state.
\label{def: env}
\end{definition}

The DET-POMDP can be viewed as a branch within the framework of POMDPs. According to the definition of POMDPs (See \ref{sec: mdps and pomdps}), the observation function maps states to observations based on the state transition function $\mathcal{T}: \mathcal{S} \times \mathcal{A} \times \mathcal{S} \rightarrow [0,1]$. If the state transition probability is deterministic (i.e., equal to $1$), it aligns with the definition of a MDP with a deterministic environment (Def. \ref{def: mdp}). In a deterministic MDP, there is no explicit observation function, and the state and observation spaces are in bijection. However, in the context of DET-POMDP (Def. \ref{def: env}), the observation is considered as a subset of the state space, which adheres to the definition of POMDPs. While the primary distinction between traditional POMDPs and the DET-POMDP problem lies in the deterministic nature of the state transition probability in the latter, both recognize that the observation space is not equivalent to the state space.

The whole process of the DET-POMDP manifests the recurrence property along with time. An agent initially starts with a state $s_t \in \mathcal{S}_b \subset \mathcal{S}$ at time-step $t \in \mathcal{T}$, after taking action $a_t \in \mathcal{A}_{s_t} \subseteq \mathcal{A}$, then the agent receives a reward $R(s_t, a_t) = r_t$ and moves to the next state $T(s_t, a_t) = s_{t+1}$. Explicitly, the state transition probability is deterministic, i.e., $P(s_{t+1} \mid s_t, a_t) = 1$ for a specific $s_{t+1}$ determined by $s_t$ and $a_t$. Iteratively, the state $s_{t+1}$ becomes the current state for the agent, and the procedure mentioned above is executed circularly along with the time series $t \in T$. Moreover, a policy of agents $\pi: \mathcal{S} \times \mathcal{A} \rightarrow [0, 1]$ gives agents guidance in action $a \in \mathcal{A}_s$ choice to a state $s \in \mathcal{S}$.

In the DET-POMDP setting, there are three types of mapping between the state space $\mathcal{S}$ and observation space $\mathcal{O}$. \textbf{(I)} The observation function $\Omega: \mathcal{S} \to \mathcal{O}$ is a \textbf{one-to-one} mapping ($\mathcal{S} = \mathcal{O}$, \cite{yu2023measuring}) when $\Omega$ is a bijection, if and only if $\forall s \in \mathcal{S}$, $\exists! o \in \mathcal{O}$, $\Omega(s)=o$. \textbf{(II)} The observation function $\Omega: \mathcal{S} \to \mathcal{O}$ is a \textbf{many-to-one} mapping, if $\mid \mathcal{S} \mid > \mid \mathcal{O} \mid$, and O is a surjection, which if and only if $\exists s \in \mathcal{S}$, $\exists o \in \mathcal{O}$, then $\Omega(s)=o$. Additionally, $\exists s_1, s_2$, consequently $\Omega(s_1) \neq \Omega(s_2)$. \textbf{(III)} The observation function $\Omega: \mathcal{S} \to \mathcal{O}$ is a \textbf{one-to-many} mapping, if $\mid \mathcal{S} \mid < \mid \mathcal{O} \mid$, and if and only if $\forall s \in \mathcal{S}$, $\exists o \in \mathcal{O}$, then $\Omega(s)=o$. While $\exists s_1, s_2 \in \mathcal{S}$ that $\Omega(s_1) \neq \Omega(s_2)$. 

This paper focuses solely on the observation function as a many-to-one mapping. In a model-free context, agents are unable to foreknow the state transition function $T$ or the observation function $\Omega$. When the observation function $\Omega$ operates as a one-to-one or one-to-many mapping, observations serve as a more effective basis than states for identifying changes in the environment. Consequently, a many-to-one mapping implies that more state information is concealed within an observation, making it difficult for agents to differentiate states based on observations.


To enhance the integration between the MDP, POMDP, graph, and automata at following up, we propose the introduction of three types of history trajectories. A trajectory typically includes states, actions, rewards, and the subsequent states. Specifically, when we record states and actions separately, we can establish a state trajectory and an action trajectory.

\begin{definition}[History trajectory]
    The agent repeatedly executes the POMDP within a finite time horizon $T$ in finite episodes $\mathcal{N}$. During processes, $\mathcal{H}$ stores all elements' history trajectory, $\mathcal{H}^a$ records all actions history trajectory, $\mathcal{H}^s$ collects all observations history trajectory. In detail, 
    \begin{itemize}
        \item A history trajectory set denotes $\mathcal{H} = {h_1, h_2, \cdots, h_k}, k > 0$.
        \item A state trajectory set represents by $\mathcal{H}^{s} = \{h_t^{s}, h_{t+1}^{s}, \cdots, h_{t+k}^{s}\}$. 
        \item A state trajectory defined as $\forall h^{s} \in \mathcal{H}^s,  h^{s} = (s_t, s_{t+1}, \cdots, s_{t+n}), n \in \mathcal{N}$.
        \item A action trajectory set represents by $\mathcal{H}^{a} = \{h_t^{a}, h_{t+1}^{a}, \cdots, h_{t+k}^{a}\}$
        \item A action trajectory defined as $\forall h^{a} \in \mathcal{H}^a, h^{a} = (a_t, a_{t+1}, \cdots, a_{t+n}), n \in \mathcal{N}$.
    \end{itemize}
\end{definition}

\subsection{The Harm of DET-POMDP}

\textbf{Subsequently, the issue of DET-POMDP exemplifies a phenomenon called Cognitive Fog and analyzes the harm that this phenomenon brings to decision-making.}

In order to investigate the risks associated with DET-POMDP, this subsection focuses on examining the lower bound of the Q-value (see \ref{sec: mdps and pomdps}) variance in the context of the DET-POMDP problem.

\begin{theorem}
    In a DET-POMDP problem $\mathcal{F}$, if there exists a state $s_i \in \mathcal{S}$ with $2 \leq i < n$, where $n$ is a positive integer, and all $s_i$ share the same state information denoted by $\Omega: s_i \rightarrow o$, where $o \in \mathcal{O}$, then the real Q-values are represented as $Q(s_i, a)$, and the observed Q-values are denoted as $\hat{Q}(s_i, a)$ for any action $a \in \mathcal{A}$. The variance (accumulated bias) of the Q-values between $Q(s_i, a)$ and $\hat{Q}(s_i, a)$ is calculated as $\frac{1}{n} \sum^n_{i=1} \left[ \left(-Q(s_i, a) + \frac{1}{n} \sum^n_{j=1} Q(s_j, a) \right)^2 \right]$.
\label{theorem: variance}
\end{theorem}

See Appendix \ref{proof: theorem: variance} for proofs.

To gain a deeper understanding of the variance in Q-values, we introduce a term called the "Cognitive Fog" phenomenon. This phenomenon arises when a state, denoted as $s$, is influenced by DET-POMDP in certain instances. Cognitive Fog refers to the situation where the agent receives both positive and negative rewards after executing the same action $a$. Consequently, these opposing rewards offset each other in the Q-value associated with the state-action pair $\langle s, a \rangle$. This occurrence introduces inaccuracies in the Q-values, ultimately resulting in decision bias.

By considering Thm. \ref{theorem: variance} in conjunction with an example, we can further illustrate the Cognitive Fog phenomenon. Suppose the variance of the actual Q-values is zero, meaning that all the real Q-values are equal to each other: $Q(s_i, a) = \frac{1}{n} \sum^n_i (Q(s_i, a))$. In this scenario, the error term $\operatorname(e)$ is equal to zero. Consequently, even if DET-POMDP is present, the Cognitive Fog phenomenon does not occur because all the real Q-values are initially identical.

Conversely, in an extreme situation where a real Q-value does exist, the variance of the error term $\operatorname(e)$ tends towards infinity. This indicates that the real Q-value experiences the Cognitive Fog phenomenon.

\subsubsection{Objective}

The objective of the DET-POMDP problem is the same as MDP and POMDP: finding an optimal policy $\pi^*$ that makes the cumulative reward maximized: 

\begin{equation}
    \pi^* = \arg\max_{\pi} \mathbb{E} \left[ \sum_{t=0}^{T} R(s_t, a_t) \mid \pi \right],
\end{equation} where $s_t$ denotes the state at time $t$, $a_t \sim \pi(o_t)$ represents the action $a_t$ chosen according to policy $\pi$ based on the current observation $o_t$, and $R(s_t, a_t)$ is the reward received immediately after executing action $a_t$ in state $s_t$.

\subsection{Bionicist Process} \label{sec: Biotechnology}

\subsubsection{Biotechnology}

\begin{figure}[h]
    \centering
    \includegraphics[width=\textwidth]{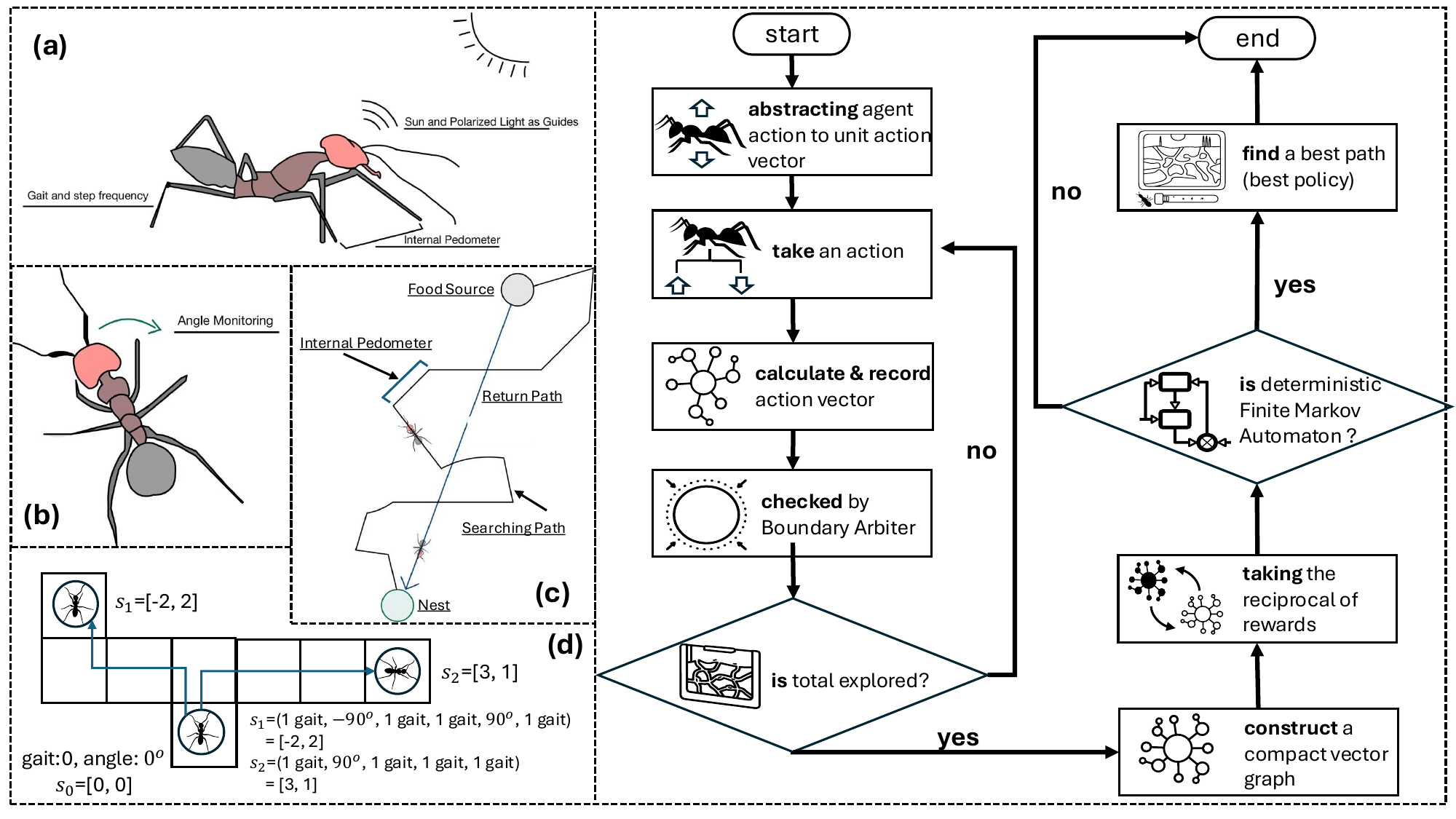}
    \label{fig: desert ant binoic}
    \caption{Schematic Diagrams and Flowchart of the Desert Ant's Navigational Biometrics and BIOMAP. (a) is a schematic diagram of the biometrics of the desert ant's navigational ability; (b) is a schematic diagram of the desert ant's angle monitoring; (c) is a schematic diagram of the desert ant's path integration; (d) is an example of the bionic algorithm; and (e) is a flowchart of the BIOMAP algorithm.}
\label{fig: desert ant bionic features}
\end{figure}

Desert ants, specifically Cataglyphis bicolor \cite{schmid1984individually}, are renowned for their exceptional navigational abilities in extreme environments. These ants possess a diverse range of bionic features that enhance their navigation skills. Fig. \ref{fig: desert ant bionic features} (a-c) visually presents some of these remarkable bionic adaptations observed in desert ants.

\paragraph{Path Integration}

Path integration \cite{muller1988path}, as described by Müller and Wehner (1988), plays a crucial role in navigation. Desert ants, when foraging for food, carefully monitor the turning angles and distances of their steps. By employing continuous path integration, they can accurately determine a direct path back to their nest, even in complex and dynamic environments.

\paragraph{Internal Pedometer}

Desert ants employ an internal pedometer mechanism to estimate walking distances, as demonstrated by \cite{wittlinger2007desert} and \cite{wehner2008desert}. However, it should be noted that the accuracy of distance measurement is influenced by factors such as gait and step frequency. Consequently, each ant possesses a unique internal pedometer that is specifically adapted to its own body shape.

\paragraph{Angle Monitoring}

Ants have the remarkable ability to monitor the angles of their turning movements during their journeys, as described by \cite{wittlinger2007desert}. Whenever an ant changes its direction, it effectively records the angle of this change, allowing it to continuously track its rotational position relative to its initial starting point. This unique navigation strategy enables ants to maintain a precise sense of orientation throughout their travels.

\paragraph{Sun and Polarized Light as Guides}

Desert ants rely on the position of the Sun as a reliable global reference point for maintaining their direction, as explained by Wehner et al. \cite{wehner2003desert}. The Sun serves as a stable reference mark that aids them in correcting their orientation. Even in cloudy or hazy conditions, when the position of the Sun may not be visible, desert ants utilize the polarization pattern of skylight (as measured on 26 August 1999 in Tunisia) to determine the direction. This celestial compass, as it is commonly referred to, enables desert ants to navigate accurately in their environment.

\paragraph{Memory and Learning}

Desert ants augment their navigational capabilities through the utilization of memory and learning mechanisms. They possess the ability to retain and recall specific routes and landmarks, enabling them to navigate more efficiently in subsequent journeys.

\subsubsection{Desert Ant Feature Extraction} \label{sec: bionic}

We abstract the bionic characteristics of Desert Ants into algorithmic elements in this subsection. Desert ants exhibit variations in gait and step frequency, which are captured through their internal pedometer that counts the number of steps taken. Additionally, they continuously monitor changes in direction angles. To replicate these features, we construct an action unit to simulate the ant's gait and an action vector to mimic the internal pedometer and angle monitoring. The action vector includes information regarding the direction and stride of each action. To create an action unit, we rely on human expertise and experience. For instance, we can abstract left and right actions into action vectors $[-1, 0]$ and $[1, 0]$ respectively, in a two-dimensional plane. We define the action unit vector accordingly.

\begin{definition}[Action Unit Vector]
     Let $\vec{a}$ be an action unit vector in the action space $\mathcal{A}$. $\nexists \vec{a^\prime}$, $\vec{0} <  \vec{a^\prime} < \vec{a}$.
\label{def: action-unit-vector}
\end{definition}

The memory and learning abilities of Desert Ants are closely associated with the graph constructed by the action unit vector, which we refer to as the "Compact Vector Graph" $\vec{G}$. In this graph, each vertex represents a cumulative action vector. For instance, if the agent moves to the right once from $[0, 0]$, the subsequent action vector would be $[1, 0]$. If the agent continues moving right from $[1, 0]$, the next action vectors would be $[2, 0]$. These three vertices, $[0, 0]$, $[1, 0]$, and $[2, 0]$, are part of $\vec{G}$, with the edges representing the action unit vectors taken by the agent.

The Compact Vector Graph $\vec{G}$ is referred to as "compact" because all the edges correspond to action unit vectors. According to Def. \ref{def: action-unit-vector}, an action unit vector represents the minimum step taken along a specific direction. Consequently, it is generally not possible to add additional vertices to the graph once the Compact Vector Graph is constructed.

\begin{definition}[Compact Vector Graph]
    Let a multigraph $\vec{G}$ be a compact Vector Graph if $\forall e \in E(\vec{G})$, $\mid \mathcal{W}(e) \mid = \vec{a}$, where $\vec{a}$ is an action unit vector.
\label{def: compact-vector-graph}
\end{definition}

In accordance with path integration, the return path from the food source to the nest corresponds to the vector connecting the source and target vertices. The length of this vector can be calculated using the Euclidean norm. However, determining the boundary of the Compact Vector Graph $\vec{G}$ poses a significant challenge. Failure to identify the boundary would result in an infinitely expanding size of the Compact Vector Graph. This challenge is thoroughly addressed in Sec. \ref{sec: boundary}. To leverage the characteristics observed in Desert Ants, we utilize environmental observations to assist in identifying the boundary, drawing inspiration from their celestial compass ability.

Similarly, various other animals exhibit similar behaviors or abilities related to path integration, which further supports the notion of its prevalence in nature. For instance, young homing pigeons possess navigation mechanisms that rely on path integration \cite{balda1998animal, jorge2008loft, jorge2006strategies}. Bees utilize a celestial-cue-based visual compass and an optic-flow-based visual odometer to achieve path integration abilities \cite{stone2017anatomically}. Additionally, swimming rats have been shown to possess path integration abilities through water maze experiments \cite{benhamou1997path}. These examples highlight the wide occurrence of path integration across different species in the animal kingdom.

Fig. \ref{fig: desert ant bionic features} (d) also shows the cumulative number of gaits and steering angles of desert ants simulated by the BIOMAP algorithm from the origin $[0,0]$ to $[-2,2]$ and $[3,1]$. In this figure, the desert ant does not utilize any information from the environment, but relies solely on its own historical gait and steering angle trajectories to define the goal states of $s_1$ and $s_2$.

\subsection{MDP-Graph-Automaton Framework} \label{sec: framework}

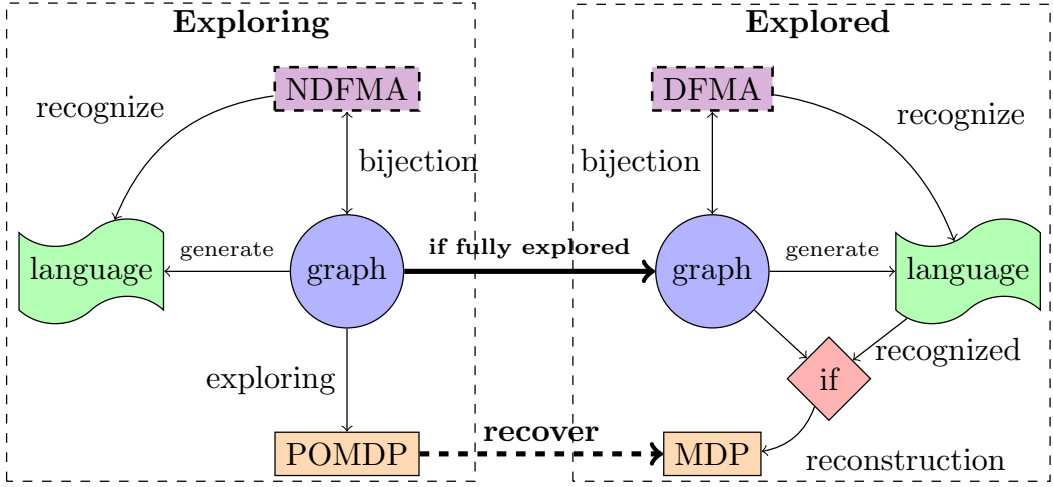
\begin{figure}[!]
  \centering
  \resizebox{0.9\textwidth}{!}{
    \begin{tikzpicture}[node distance=2cm, auto]
      \node[draw, circle, fill=blue!30] (graph1) {\textcolor{black}{graph}};
      \node[draw, circle, right of=graph1, node distance=4cm, fill=blue!30] (Graph) {\textcolor{black}{graph}};
      \node[draw, shape=tape, tape bend height=0.3cm, node distance=2.8cm, left of=graph1, fill=green!30] (language) {\textcolor{black}{language}};
      \node[draw, shape=tape, tape bend height=0.3cm, node distance=2.8cm, right of=Graph, fill=green!30] (Language) {\textcolor{black}{language}};
      \node[draw, rectangle, below of=graph1, fill=orange!30] (pomdp) {\textcolor{black}{POMDP}};
      \node[draw, thick, dashed, above of=graph1, fill=violet!30] (ndfma) {\textcolor{black}{NDFMA}};
      \node[draw, thick, dashed, above of=Graph, fill=violet!30] (dfma) {\textcolor{black}{DFMA}};
      \node[draw, rectangle, below of=Graph, fill=orange!30] (mdp) {\textcolor{black}{MDP}};
      \node[draw, diamond, below right=0.5cm and 0.6cm of Graph, fill=red!30] (if) {\textcolor{black}{if}};
    
      \draw[->] (graph1) -- (pomdp) node[midway, left] {exploring};
      \draw[<->] (graph1) -- (ndfma) node[midway, right] {bijection};
      \draw[<->] (Graph) -- (dfma) node[midway, left] {bijection};
      \draw[->] (Graph) -- (if) node[pos=0.8, left] {};
      \draw[->] (Language) -- (if) node[pos=0.8, right] {recognized};
      \draw[->] (if) to[bend left] node[pos=0.5, below right] {reconstruction} (mdp);
      \draw[->, ultra thick] (graph1) -- (Graph) node[midway, above] {\scriptsize \textbf{if fully explored}};
      \draw[->] (graph1) -- (language) node[midway, above] {\scriptsize generate};
      \draw[->] (Graph) -- (Language) node[midway, above] {\scriptsize generate};
    
      \draw[->] (ndfma) to[bend right] node[pos=0.5, above left] {recognize} (language);
      \draw[->] (dfma) to[bend left] node[pos=0.5, above right] {recognize} (Language);

      \draw[->, dashed, ultra thick] (pomdp) to[left] node[pos=0.5, above] {\textbf{recover}} (mdp);
      
      \draw[dashed] ([shift={(-3.1cm,-2.3cm)}]graph1.west) rectangle ([shift={(5cm,2.5cm)}]language.north west);  
      
      \draw[dashed] ([shift={(-0.9cm,-2.3cm)}]Graph.west) rectangle ([shift={(0.1cm,2.5cm)}]Language.north east);

      \node[above right, xshift=-35pt, yshift=5pt] at (ndfma.north west) {\textbf{Exploring}};
      \node[above left, xshift=40pt, yshift=5pt] at (dfma.north east) {\textbf{Explored}};
      
    \end{tikzpicture}%
  }
  \caption{Structure of MDP-Graph-Automaton.}
  \label{fig: construction}
\end{figure}

The aim of this subsection is to establish a connection between MDPs, graphs, and automata. As depicted in Fig. \ref{fig: construction}, the exploration phase involves the interaction between the graph and the POMDP. The graph records the action vector while simultaneously maintaining a mutually isomorphic relationship with the Non-deterministic Finite Markov Automaton (NDFMA). The NDFMA can recognize the language generated by the trajectories of the graph.

Upon completion of the exploration phase, the historical trajectory generation language of the graph can be recognized by the Deterministic Finite Markov Automaton (DFMA), which is in mutual isomorphism with the graph. This implies that the state and unit action vectors recorded on the graph can be transformed into fully observable MDP. Subsequently, the shortest path algorithm can be employed to identify the optimal path (best policy) on the graph.

Hence, the MDP-Graph-Automaton framework at the core of our approach involves the conversion of a POMDP problem into an MDP problem, followed by the solution of the MDP problem.

First, we introduce the basics of automaton.

\subsubsection{MDP-Graph Relation}

Def. \ref{def:GMDP-graph} conveys the relation between graph and MDP.

\begin{definition}[MDP with graph representation]
A Markov Decision Process with graph representation $\mathcal{M}G$ is a multigraph. It is formalized by the tuple $\langle \mathcal{V}, E \rangle$. $\mathcal{M}G$ is constructed through deterministic Markov Decision Process $\mathcal{M}= \langle \mathcal{S}, \mathcal{A}, \mathcal{T}, \mathcal{R}, \gamma \rangle$. Corresponding to $\mathcal{M}$, $\mathcal{M}G$ satisfied the bijective mapping relations as follow: (I) $\mathcal{V} \leftrightarrow \mathcal{S}$; (II) $\mathcal{W}(E) \leftrightarrow \mathcal{R}$, where $\mathcal{W}(E)$ is the weight of edges; (III) $\mathbb{A}(E) \leftrightarrow \mathcal{A})$, where $\mathbb{A}(E)$ is the action attribute of edges.
\label{def:GMDP-graph}
\end{definition}

\subsubsection{MDP-Automaton Relation}

To enhance the representation of automata in the context of MDPs, we introduce the concept of a Markov Automaton. The Markov Automaton can be classified as either a Deterministic Finite Markov Automaton (DFMA) or a Non-Deterministic Finite Markov Automaton (NDFMA), as defined in Def. \ref{def:automaton}. This addition allows us to strengthen the incorporation of automata in the MDP framework.

\begin{definition}[Dual relation]
Let $\mathbb{M} =\langle \mathcal{S}, \mathcal{A}, \mathcal{T}, \mathcal{R}, \gamma \rangle$ be a MDP and $\mathcal{M}_{\text{Markov}} = \langle Q, \Sigma, \delta, q_0, F \rangle$ be a Markov automaton. The dual relation between $\mathbb{M}$ and $\mathcal{M}_{\text{Markov}}$, denoted as $\mathbb{M} \sim M_{\text{Markov}}$, exists if the following bijective mappings hold: (I) $\mathcal{S} \leftrightarrow Q$; (II) $\mathcal{A} \leftrightarrow \delta$; (III) $s_0 \leftrightarrow q_0$; (IV) $s_{\text{terminal}} \leftrightarrow F$, where $s_{\text{terminal}}\in \mathcal{S}$ is the terminal states.
\label{def: dual relation}
\end{definition}

\subsubsection{MDP-Graph-Automaton Relation}

We define a language based on Def. \ref{def: transition relation}, where a word is a sequence consisting of actions, denoted as $w = (a_0, a_1, \cdots, a_n)$, with $n > 0$. The empty word is represented by $\varepsilon$. 
\textbf{Then, we aim to demonstrate that a DFMA $\mathcal{M}_{Markov}$ can effectively recognize the historical trajectory.}

\begin{lemma}
    Let a MDP $\mathbb{M} = (S, A, R, \gamma)$ and a Deterministic Finite Markov Automaton $\mathcal{M}_{Markov} = (Q, \Sigma, \rho, s_0, F)$, then the history trajectory set produced by $\mathbb{M}$: $\mathcal{H} = L_{Markov}$. The language $L_{Markov}$ recognizable by $\mathcal{M}_{Markov}$.
\label{lemma:mdp to dfma}
\end{lemma}

See Appendix \ref{proof: lemma:mdp to dfma} for proofs.

We prove the action history trajectory set $\circ h^a$ is a subset of the alphabet $\Sigma$ and exists a word $w$ equal to the $h^a$. Moreover, the transition relation $(q_0, w) \vdash (q, \varepsilon)$ exists. Thus, we can use the $L_{Markov}$ to do the Reinforcement Learning.

\subsubsection{MDP-Graph-Automaton Framework in POMDP}

According to Def. \ref{def: env}, $\mathcal{O}$ is the observation space, illustrating that an observation may be any state. Yu et al. \cite{yu2023measuring} proposed a conception of the gap between state and observation and clearly introduced that the observation has a one-to-many mapping relationship to the state. Thus, based on Def. \ref{def:automaton}, \textbf{the POMDP corresponds to NDFMA.}

\begin{lemma}
    Let a POMDP $\mathbb{P} = \langle \mathcal{S}, \mathcal{A}, \mathcal{O}, \Omega, \mathcal{R}, \mathcal{T}, \gamma \rangle$ and a Nondeterministic Finite Markov Automaton $M^\prime_{Markov} = (Q^\prime, \Sigma^\prime, \rho, s_0, F)$, then the history trajectory set produced by $M^\prime$: $\mathcal{H}^\prime = L_{Markov}^\prime$. The $L_{Markov}^\prime$ recognizable by $M_{Markov}^\prime$.
\label{lemma:pomdp to ndfma}
\end{lemma}

See Appendix \ref{proof: lemma:pomdp to ndfma} for proofs.

Therefore, based on Lem. \ref{lemma:mdp to dfma} and \ref{lemma:pomdp to ndfma}, we have Thm. \ref{theorem: deterministic = nondeterministic language}.

\begin{theorem}
    The class of languages recognized by Non-Deterministic Finite Markov Automata is equivalent to the class of languages recognized by Deterministic Finite Markov Automata.
\label{theorem: deterministic = nondeterministic language}
\end{theorem}

According to Thm. \ref{theorem: deterministic = nondeterministic language}, the language $L$ from the Fully Observable MDP $M$ is also recognizable by NDFMA $M^\prime$. We aim to reduce the variance of every pair $\langle o , a, r \rangle$ and transfer NDFMA to DFMA. \textbf{More importantly, if a language is not recognizable by DFMA, then the corresponding MDP for that language is partially observable. We want to construct a graph that corresponds to a generated language that can be recognized by DFMA.}

\subsection{Boundary Arbiter} \label{sec: boundary}

Defining boundaries for the Action Vector Graph $\vec{G}$ poses a challenge. In certain scenarios, there may be no environmental limitations that restrict the agent from taking actions indefinitely. For instance, a car can continue driving endlessly on an infinite plane, or an injection device can be continuously pushed without knowledge of the patient's condition. Consequently, if the agent lacks prior knowledge about the environment, it can easily become trapped in an unbounded problem.

To address this issue, we propose the Boundary Arbiter as a solution. The Boundary Arbiter is designed to prevent the agent from entering boundless situations by imposing constraints or limits on the actions that can be taken. By incorporating the Boundary Arbiter, we aim to provide the agent with a mechanism to navigate and operate within well-defined boundaries, mitigating the challenges associated with unbounded environments.

In Alg. \ref{alg: boundary}, the Boundary Arbiter utilizes state observations to aid in the determination of boundaries. When the agent receives the next observation $o^\prime$ that is the same as the current observation $o = o^\prime$ after taking an action $a$, indicating it may be a boundary, the Boundary Arbiter identifies this loop as the initial indication of a boundary. This is because, when the agent encounters a boundary and continues to take actions that interact with the boundary, it will remain in the same state and receive the same observation repeatedly.

The algorithm consists of two distinct parts. In the first part, a boundary tolerance $\delta \in \mathbb{Z}^+$ is set to determine whether the observation is located on the inner or outer boundary. Subsequently, all actions $a_j$ from the action space $\mathcal{A}$ are taken at the observation $o$, resulting in the corresponding next observation $o_j$.

In the second part the agent repeatedly takes the same action for $\delta - 1$ times, continuing to interact with the environment. If the next observation $o^\prime$ is different from the current observation $o$, False is returned as they obviously correspond to different states. Subsequently, for each subsequent observation $o^\prime$, all actions $a_i \in \mathcal{A}$ are taken and the corresponding next observation $o^\prime_j$ is received. If at any point $o_j \neq o^\prime_j$, False is returned as it indicates that the current observation is not the same as the next observation. Finally, if the observation $o$ passes all the checks in both parts, it is deemed to be a boundary, and a self-loop is created in the Compact Action Graph $\vec{G}$.

It is important to note that an observation has two distinct identities. One identity is obtained from the environment and is represented by a numeric value $o$. The other identity is derived from the action vector and is denoted by $\vec{o}$.

\begin{algorithm}[t!]
\caption{Boundary Arbiter Algorithm $\mathcal{B}$.}
\begin{algorithmic}[1]
    \STATE  \textbf{Input: } An DET-POMDP problem environment $\mathcal{F}$, an Compact Vector Graph $\vec{G}$, an observation $o$, an action vector $\vec{o}$ of the observation $o$, an action $a$.
    \STATE  \textbf{Output: } The observation $o$ is a boundary or not (True or False).
    \STATE  Initialize an tolerance $\delta$. Set $i=0$ and $j=0$.
    \FOR{ $a_j$ in $\mathcal{A}$}
        \STATE  Take the action $a_j$ at the observation $o$, and receive the next observation $o_j$.
    \ENDFOR
    \FOR{ $i$ in $2:\delta$}
        \STATE  Take the action $a$ at the observation $o$ (maps to $\vec{o}$), and receive the next observation $o^\prime$ (maps to $\vec{o^\prime}$).
        \IF{ $o \neq o^\prime$}
            \STATE  return False
        \ENDIF
        \FOR{ $a_j$ in $\mathcal{A}$}
        \STATE  Take the action $a_j$ at the observation $o$ (maps to $\vec{o}$), and receive the next observation $o^\prime_j$.
        \IF{ $o_j \neq o^\prime_j$}
            \STATE  return False
        \ENDIF
        \ENDFOR
        \STATE  $o = o^\prime$, $\vec{o} = \vec{o^\prime}$
    \ENDFOR
    \STATE  Build a self-loop $(\vec{o}, a, \vec{o})$ in $\vec{G}$.
    \STATE  return True
\end{algorithmic}
\label{alg: boundary}
\end{algorithm}

\subsection{BIOMAP Algorithm} \label{sec: BIOMAP}

\begin{algorithm}[!]
\caption{BIOMAP Algorithm.}
\begin{algorithmic}[1]
    \STATE  \textbf{Input: } An DET-POMDP problem environment $\mathcal{F}$
    \STATE  \textbf{Output: } The best policy $\pi^*$
    \STATE  Initialize an Compact Vector Graph $\vec{G} = \langle \mathcal{V}, E$ with vertex space $\mathcal{V} = \varnothing$, edge space $E = \varnothing$. Construct action unit vector set $\vec{\mathcal{A}}$. Set total explored degree $\Phi = 0$. Set maximum step number is $M$, and the maximum episode number is $N$.

    \WHILE{ $\Phi \neq 0$ or $n < N$}
        \STATE  Get the initial observation $o$, and insert vertex $v = \vec{s}$ 
        \STATE  Update the total explored degree $\Phi = \Phi + \Phi_{\vec{s}}$, where $\mid \Phi_{\vec{s}} \mid = \mid \vec{\mathcal{A}} \mid$.
        \STATE  $n = n + 1$.
            
        \WHILE{ $m < M$}
            \STATE  Find the action $a$ that never be taken at observation $o$, and execute $a$.
            \STATE  Receive the reward $r$ and the next observation $o^\prime$.
            \IF{ $o = \vec{o}$}
                \STATE  Use $b = \mathcal{B}(\mathcal{F}, \vec{G}, o, \vec{o}, a)$ (From Alg. \ref{alg: boundary}).
            \ENDIF
            \IF{ $b$ is False}
                \STATE  Calculate the next action vector $\vec{o^\prime} = \vec{o} + \vec{a}$.
                \STATE  Insert edge $e = (\vec{o}, \vec{o^\prime})$, where $\mathcal{W}(e) = r$ and $\mathcal{A}(e) = \vec{a}$.
                \STATE  Update the total explored degree $\Phi = \Phi + \Phi_{\vec{s^\prime}}  - 1$, where $\mid \Phi_{\vec{o^\prime}} \mid = \mid \vec{\mathcal{A}} \mid$.
                \STATE  $o = o^\prime$.
            \ENDIF
            \STATE  $m = m + 1$.
        \ENDWHILE
    \ENDWHILE

    \STATE  Construct a Finite Markov Automaton $\mathcal{M}_{\text{Markov}}$ which is isomorphism to $\vec{G}$.
    \IF{ $\mathcal{M}_{\text{Markov}}$ is not deterministic} 
        \STATE  Break
    \ENDIF
    \IF{ $\forall e \in \mathcal{E}$ that $min(\mathcal{W}(e)) < 0$}
        \STATE  $\mathcal{W}(e) = \mathcal{W}(e) + |min(\mathcal{W}(e))|$
    \ENDIF
    \STATE  $\forall e \in E(G)$ that $\mathcal{W}(e) = max(\mathcal{W}(e)) - \mathcal{W}(e)$
    \IF{ exist negative self-loop or cycle in $\vec{G}$}
        \STATE  Break
    \ENDIF
    \STATE  Use the Dijkstra algorithm to find the shortest from start to terminal.
    \STATE  Change the shortest path to a policy and return the policy.
    
\end{algorithmic}
\label{alg: biomap}
\end{algorithm}

The flowchart of BIOMAP is given in Fig. \ref{fig: desert ant bionic features} (e) and the flowchart shows the main steps of BIOMAP. In Alg. \ref{alg: biomap}, we present the pseudocode for the BIOMAP algorithm. The input is a DET-POMDP problem environment $\mathcal{F}$, and the output is the optimal policy $\pi^*$.


During the initialization phase, we create an empty Compact Vector Graph $\vec{G}$ (as defined in Def. \ref{def: compact-vector-graph}) and abstract the agent's action space $\mathcal{A}$ into the action unit vector space $\vec{\mathcal{A}}$. Additionally, we set the exploration degree $\Phi$ of all vertices to zero. Since the DET-POMDP problem is deterministic (as per Def. \ref{def: env}), our exploration strategy is to take all action unit vectors for every vertex, ensuring that the agent fully explores the environment. Finally, we specify the maximum number of episodes $N$ (outer loop) and steps $M$ (inner loop).

We divide the algorithm into four phases. The first phase is the construction exploration phase, which stops either when the environment is fully explored by the agent or when the maximum interaction limit is reached. The environment initializes the observation $o$ and inserts the initial action vector $\vec{o}$ into $\vec{G}$. Since no action is taken at this vertex, the exploration degree is set to $\mid \mathcal{A} \mid$, and the total exploration degree is updated as $\Phi = \Phi + \Phi_{\vec{s}}$. After the environment initialization, the agent selects an action $a$ that will not be taken at observation $o$. The environment then provides the reward $r$ and the next observation $o^\prime$ (line 10). Subsequently, Alg. \ref{alg: boundary} is used to determine if the current observation is a boundary. If it is a boundary, the agent remains in the same observation. Otherwise, the action vector of observation $\vec{o^\prime}$ is obtained by adding the action vector unit $\vec{a}$, which is derived from the action $a$, to the previous action vector $\vec{o}$, resulting in $\vec{o^\prime} = \vec{o} + \vec{a}$. An edge $e = (\vec{o}, \vec{o^\prime})$ is inserted into the graph with the weight $\mathcal{W}(e) = r$ and attribute $\mathcal{A}(e) = \vec{a}$. At this point, there are two vertices, and the original vertex has taken an action $a$, thus updating the total exploration degree as $\Phi = \Phi + \Phi_{\vec{s^\prime}} - 1$. Finally, the current observation is set to the next observation $o = o^\prime$, and the first phase is iteratively executed until the stopping condition is reached.

The second phase involves determining whether $\vec{G}$ is deterministic. According to Thm. \ref{theorem: deterministic = nondeterministic language}, if the isomorphism of $\vec{G}$, which is represented by the Markov Automaton, is deterministic, then $\vec{G}$ is also deterministic. The construction of the Markov Automaton is defined in Def. \ref{def: dual relation}.

In the third phase, we transform the maximum path-finding problem (aiming to find a policy with maximum cumulative reward) into a minimum path-finding problem. This is necessary because the maximum path-finding problem is known to be NP-hard \cite{karp2010reducibility}. To achieve this transformation, We make the weight of all edges positive and reverse the order of large and small numbers. We then check whether there is negative self-loops or cycles in $\vec{G}$. If such loops or cycles exist, the concept of the shortest path becomes meaningless \cite{cormen2022introduction}.

The fourth phase involves using the Dijkstra algorithm \cite{dijkstra2022note} to find the shortest path and convert it into a policy. Our objective is to find the shortest path since we reverse weight numbers in the third phase. Once we obtain the shortest path, which consists of vertex sequences $(\vec{s_1}, \vec{s_2}, \cdots, \vec{s_n})$, we determine the lowest-weight edge between any pair of consecutive vertices. This enables us to construct the optimal policy $\pi^*(s) = a$, where $s$ and $a$ belong to the vertices of the shortest path.

\subsection{Time Complexity Analysis} \label{sec: time complexity}

For Alg. \ref{alg: boundary}, setting variables value is a constant time complexity $O(1)$. The $\delta - 1$ times outer loop has a time complexity of $O(\delta)$, and the $\mid \mathcal{A} \mid$ times inner loop has a time complexity of $O(\mid \mathcal{A} \mid)$. Combined with all taking action operations, the time complexity of taking action is $O(\delta \cdot \mid \mathcal{A} \mid)$. For the judgment operations, even if the time complexity of these operations is $O(1)$, they are in the episode, so we consider the time complexity of judgment operation to be $O(\mid \delta \mid)$ and $O(\delta \cdot \mid \mathcal{A} \mid)$. The rest operation has a time complexity of $O(1)$. Summary above all, we get the total time complexity of Alg. \ref{alg: boundary} is $O(\delta)+O(\mid \mathcal{A} \mid)+O(\delta \cdot \mid \mathcal{A} \mid)+O(\delta)+O(\delta \cdot \mid \mathcal{A} \mid)$. Simplifying the time complexity to be

\begin{equation}
    O(\delta \cdot \mid \mathcal{A} \mid).
\end{equation}

For Alg. \ref{alg: biomap}, the outer loop at most has $N$ times, so the time complexity of the outer loop is $O(N)$. Because the maximum number of episodes is $M$ for the inner loop, the time complexity of the inner loop is $O(M)$. In the inner loop, the interaction operation has a time complexity of $O(1)$, and the second judgment operation also has a time complexity of $O(1)$. The $B$ function comes from Alg. \ref{alg: boundary}, so for the first judgment operation, the time complexity is $O(\delta \cdot \mid \mathcal{A} \mid)$ because the rest operation in the out loop is $O(1)$. Therefore, the time complexity of the first phase is $O(N \cdot M \cdot \delta \cdot \mid \mathcal{A} \mid)$. In the second phase, because the graph and automaton are isomorphisms, thus we think this is a copy operation in which the time complexity is $O(\mathcal{V} + E)$. The deterministic checking operation also traverses all states and alphabets of the Markov Automaton. So the time complexity of the second phase is $O(2 \cdot (\mathcal{V} + E))$, and simplifying it as $O(\mathcal{V} + E)$. In the third phase, the updating edge weight operation has the time complexity of $O(E)$. The negative self-loop or cycle checking operation also traverses all vertices and edges. Thus, the third phase has the time complexity of $O(\mathcal{V} + 2 \cdot E)$, and we simplify it as $O(\mathcal{V} + E)$. The fourth phase is the Dijkstra algorithm, which has the time complexity of $O(\mathcal{V}^2)$. In summary, Alg. \ref{alg: biomap} has the time complexity of

\begin{equation}
    \begin{aligned}
     & O(N \cdot M \cdot \delta \cdot \mid \mathcal{A} \mid) + O(\mathcal{V} + E) + O(\mathcal{V} + E) + O(\mathcal{V}^2)\\
     & = O(N \cdot M \cdot \delta \cdot \mid \mathcal{A} \mid + \mathcal{V}^2 + E)
    \end{aligned}
\end{equation}

Combining the time complexity of Alg. \ref{alg: boundary} and \ref{alg: biomap}, we have the total time complexity of 

\begin{equation}
    O(N \cdot M \cdot \delta \cdot \mid \mathcal{A} \mid + \mathcal{V}^2 + E)
\label{equ: time complexity}
\end{equation}

\section{Experiment} \label{sec: experiment}

\subsection{Experiment Design}

To evaluate the effectiveness of the BIOMAP algorithm in DET-POMDP problem, we design a scenario using the Cliff Walking \cite{sutton2018reinforcement} environment as the foundation of our simulator. The choice of Cliff Walking is motivated by its classification as a grid world simulator, which allows for the abstraction of various real-world scenarios. Grid worlds have been extensively utilized in different domains. For instance, Thrun \cite{thrun1994learning} applied grid worlds to robotics navigation problems, Wiering \cite{wiering2000multi} demonstrated how urban traffic control can be simplified to a grid world setting, and Bechar et al. \cite{bechar2016agricultural} introduced the concept of precision agriculture where farmlands are divided into grids to construct an intelligent agriculture system for resource allocation. Therefore, the grid world serves as an exemplary model for abstracting real-life situations.

\begin{figure}[h!]
    \centering
    \includegraphics[width=\textwidth]{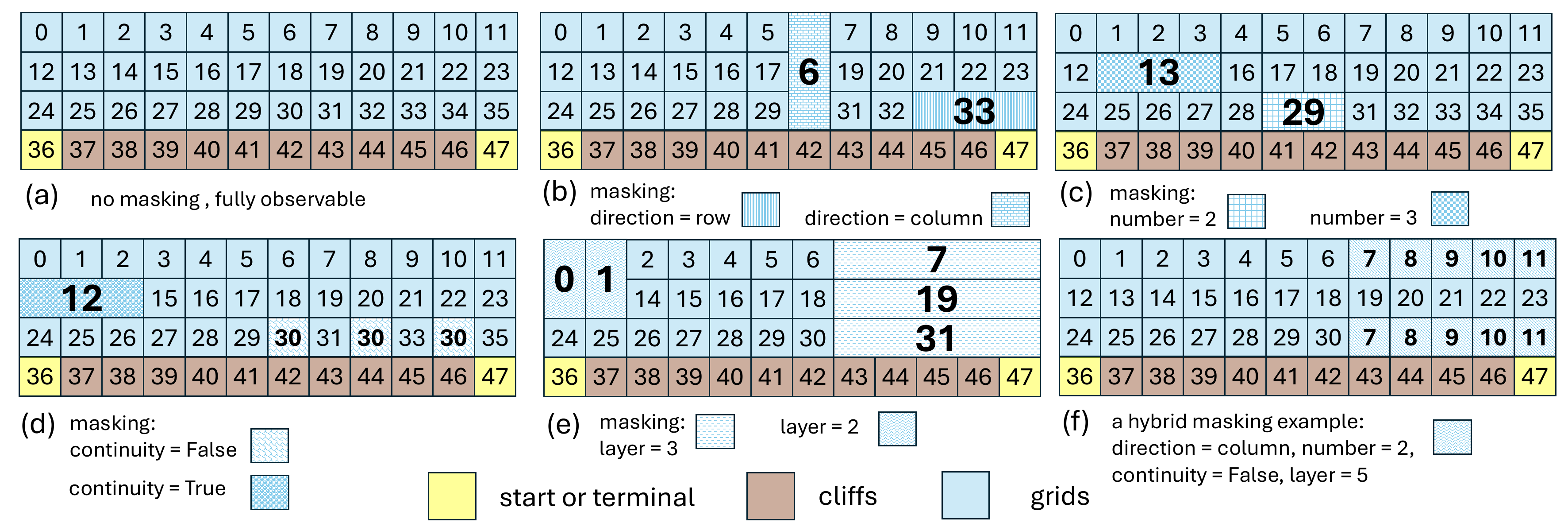}
    \caption{This figure illustrates the design of the Masking Cliff Walking experiment, showcasing different settings of observability. In setting (a), the environment is fully observable, denoted by the absence of any masking. The agent has complete knowledge of the grid configuration at all times. In settings (b) to (f), the environment becomes partially observable, introducing various forms of masking to limit the agent's visibility of the grid world. Setting (b) demonstrates different masking directions; Setting (c) exhibits different masking numbers; Setting (d) shows a combination of continuous and discrete masking; Setting (e) displays different masking layers; Setting (f) presents a hybrid setting that combines elements from settings (b) to (e).}
\label{fig: experiment}
\end{figure}

The Masking Cliff Walking environment is similar to the traditional version and consists of a total of $48$ grids. In Fig. \ref{fig: experiment}, grids $0$ to $35$ are classified as normal grids, grid $36$ serves as the starting point, and grid $47$ is the terminal point. Grids $37$ to $46$ are designated as cliffs. When the agent steps on a normal grid, it receives a reward feedback of $-1$. If the agent mistakenly steps into a cliff, it incurs a penalty of $-100$, and the environment immediately terminates. On the other hand, when the agent successfully reaches the terminal grid, it receives a reward of $+10$.

The objective of this experiment is to identify the optimal policy that guides the agent to reach the terminal grid while maximizing its cumulative rewards. It is evident in Fig. \ref{fig: experiment} that the best path to achieve this objective is through the sequence of grids $(36, 24, 25, 26, 27, 28, 29, 30, 31, 32, 33, 34, 35, 47)$, which leads to a maximum cumulative reward of $-2$.

\begin{table}[htbp]
    \centering
    \begin{threeparttable}
    \caption{Comparison rewards in terms of descriptive statistics analysis. ($N=84$)} 
    \begin{tabular}{|c|c|c|c|c|c|c|c|c|}
    \hline
    \tiny ID & \tiny Algorithm & \tiny Online/Offline & \tiny Model-based or -free & \tiny Mean & \tiny Maximum & \tiny Minimum & \tiny Variance & \tiny time cost\\
    \hline 
    \tiny $1$ & \tiny QMDP \cite{littman1995learning} & \tiny Offline & \tiny Model-based & \tiny $-2.00$ & \tiny $-2$ & \tiny $-2$ & \tiny $0.00$ & \tiny $<0.01$ \\
    \tiny $2$ & \tiny FIB \cite{kochenderfer2022algorithms} & \tiny Offline & \tiny Model-based & \tiny $-4.22$ & \tiny $-2$ & \tiny $-50$ & \tiny $103.02$  & \tiny $587.17$ \\
    \tiny $3$ & \tiny SARSOP \cite{kurniawati2009sarsop} & \tiny Offline & \tiny Model-based & \tiny $-2.00$ & \tiny $-2$ & \tiny $-2$ & \tiny $0.00$ & \tiny $0.33$\\
    \tiny $4$ & \tiny PO-UCT \cite{kocsis2006bandit} & \tiny Offline & \tiny Model-based & \tiny $-49.66$ & \tiny $-28$ & \tiny $-50$ & \tiny $7.45$ & \tiny $178.67$\\
    \tiny $5$ & \tiny AdaOPS \cite{wu2021adaptive} & \tiny Online & \tiny Model-based & \tiny $-47.80$ & \tiny $-14$ & \tiny $-50$ & \tiny $54.38$ & \tiny $1443.17$\\
    \tiny $6$ & \tiny POMCPOW \cite{sunberg2018online} & \tiny Online & \tiny Model-based & \tiny $-50.00$ & \tiny $-50$ & \tiny $-50$ & \tiny $0.00$ & \tiny $483.00$ \\
    \tiny $7$ & \tiny POMDPSolver & \tiny Offline & \tiny Model-based & \tiny $-2.00$ & \tiny $-2$ & \tiny $-2$ & \tiny $0.00$ & \tiny $>4277.33$\\
    \tiny $8$ & \tiny BIOMAP & \tiny Offline & \tiny Model-free & \tiny $-2.00$ & \tiny $-2$ & \tiny $-2$ & \tiny $0.00$ & \tiny $1.67$ \\
    \hline
    \end{tabular}
\label{tab: descriptive}
\end{threeparttable}
\end{table}

\begin{table}[h!]
\centering
\begin{threeparttable}
    \caption{ANOVA Analysis Table for Cumulative Rewards.}
    \begin{tabular}{|c|c|c|c|c|c|c|c|}
    \hline
    \scriptsize ID & \scriptsize Dep. Vari. &  \scriptsize Indep. Vari. & \scriptsize Reference &  \scriptsize P-value & \scriptsize Signif.$^1$ & \scriptsize Max. Mean \\
    \hline
    \scriptsize $1$ & \scriptsize Rewards & \scriptsize Algorithms & - & \scriptsize $<2e-16$ & \scriptsize $***$ & \scriptsize BIOMAP/SARSOP/POMDPSolver/QMDP \\
    \scriptsize $2$ & \scriptsize Rewards & \scriptsize Masking directions & \scriptsize Fig. \ref{fig: experiment} (b) & \scriptsize $0.809$ & $-$  & - \\
    \scriptsize $3$ & \scriptsize Rewards & \scriptsize Masking numbers & \scriptsize Fig. \ref{fig: experiment} (c) & \scriptsize $1$ & $-$ & -  \\
    \scriptsize $4$ & \scriptsize Rewards & \scriptsize Masking continuity & \scriptsize Fig. \ref{fig: experiment} (d) & \scriptsize $0.83$ & $-$ & - \\
    \scriptsize $5$ & \scriptsize Rewards & \scriptsize Masking layers & \scriptsize Fig. \ref{fig: experiment} (e) & \scriptsize $1$ & $-$ & - \\
    \scriptsize $6$ & \scriptsize Rewards & \scriptsize is BIOMAP & - & \scriptsize $7.29e-10$ & \scriptsize $***$ & \scriptsize Yes$^2$ \\
    \scriptsize $7$ & \scriptsize Rewards & \scriptsize is Offline & - & \scriptsize $<2e-16$ & \scriptsize $***$ & \scriptsize Yes$^3$ \\
    \hline
    \end{tabular}
    \begin{tablenotes} 
    \item[1] \scriptsize Significance codes:  $\mid 0 \sim 0.001: *** $ $\mid 0.001 \sim 0.01: **$  $\mid 0.01 \sim 0.05: *$ $\mid 0.05 \sim 0.1: .$ $\mid 0.1 \sim 1: - \mid$
    \item[2] \scriptsize "Yes" means that the mean value of the rewards of the BIOMAP algorithm is greater than that of the non-BIOMAP algorithm.
    \item[3] \scriptsize "Yes" means that the reward mean of the Offline algorithm is greater than that of the non-offline (online) algorithm.
    \end{tablenotes}
\label{tab: anova}
\end{threeparttable}
\end{table}

In order to align the Cliff Walking environment with the definition of DET-POMDP (Def. \ref{def: env}), we have implemented a \textbf{masking mechanism} that ensures consistency and uniformity of state information within the masked regions. This mechanism allows the environment to respond in the same way to the same state of information for all grids covered by masking. The \textbf{masking direction} can be either horizontal or vertical, with all maskings within a specific setting maintaining the same direction. Both \textbf{continuous and discrete masking} are allowed, but not simultaneously within a single setting. Discrete masking includes a non-masking grid in the middle with intervals between two masked grids, while continuous masking indicates a continuous region without any non-masking grids within the masked area. \textbf{The size of the masking} determines the level of partial observability, with maximum lengths specified for each type of masking. The maximum step number for each step is set at $50$ to limit the agent's exploration and decision-making within the environment. These design considerations ensure that the Masking Cliff Walking experiment meets the criteria of DET-POMDP definitions while providing a variety of partially observable scenarios for evaluating the performance of the BIOMAP algorithm.

To investigate the impact of the masking mechanism on algorithm performance, we have developed various settings that differ in four dimensions: masking direction, continuity, size of masking, and count of masking layers. These multidimensional changes ensure a diverse range of settings, thereby avoiding high randomness in the results. 
Fig. \ref{fig: experiment} provides a visual representation of the experiment scenario, with sub-figure (f) illustrating an example that incorporates all four dimensions.

In this example, ten regular grids $(7, 31, 8, 32, 9, 33, 10, 34, 11, 35)$ are covered by masking. The masking layers are set to five, ensuring that each layer has a distinct masking observation value. For instance, grids $7$ and $31$ belong to the same layer, so the environment returns observation $7$ regardless of whether the agent is at grid $7$ or $31$. The masking direction is set to columns, meaning that each layer of columns follows the same setting design. The continuity is set to false, resulting in discrete maskings where no maskings cover the middle grid between any two maskings at any layer. The masking number is set to $2$, indicating that there are two maskings in each layer. Regardless of the masking setting, the masking always starts from grid $35$. These comprehensive settings allow us to systematically explore the impact of different masking dimensions on algorithm performance in the Masking Cliff Walking experiment.

\subsection{Comparison Algorithms}

To ensure consistency in experimental conditions, we have implemented the BIOMAP algorithm and Mask Cliff Walking environment using Julia-1.9.3 \cite{bezanson2017julia}. The comparative algorithms mentioned in Sec. \ref{sec: related works} are sourced from POMDPs.jl \cite{egorov2017pomdps}. Statistical analysis and visualization of the resulting data are performed using R-4.2.0 \cite{team2013r}. All the comparison algorithms are executed on Julia-1.9.3 using the official functions and methods provided by POMDPs.jl. The experimental setup utilized a Macbook Pro equipped with an Apple M1 chip, 8GB of RAM, and macOS 14.4.1 (23E224) as the hardware environment for all the experiments conducted.

We have selected several comparison algorithms within the POMDPs.jl ecosystem. These include QMDP \cite{littman1995learning} (QMDP.jl), FIB \cite{kochenderfer2022algorithms} (FIB.jl), SARSOP \cite{kurniawati2009sarsop} (SARSOP.jl), PO-UCT \cite{silver2010monte} \cite{kocsis2006bandit} (BasicPOMCP.jl), AdaOPS \cite{wu2021adaptive} (AdaOPS.jl), POMCPOW \cite{sunberg2018online} (POMCPOW.jl), and POMDPSolver (POMDPSolve.jl). Notably, POMDPSolve.jl provides a diverse range of algorithms, such as Enumeration \cite{sondik1971optimal}, Two Pass \cite{sondik1971optimal}, Linear Support \cite{cheng1988scaling}, Witness \cite{littman1994witness}, Incremental Pruning \cite{cassandra1997incremental}, and Finite Grid \cite{cassandra2004tony}.

However, there are several reasons why certain algorithms cannot be compared in our study. Firstly, the AEMS algorithm (AEMS.jl) is not applicable to the DET-POMDP scenarios due to the observation function limitations in Mask Cliff Walking. Secondly, the Belief Grid Value episode algorithm (BeliefGridValueepisode.jl) and the Particle Filter Trees with Double Progressive Widening algorithm (ParticleFilterTrees.jl) lack the necessary updater module, making them incompatible for comparison. Lastly, the MCVI algorithm (MCVI.jl) faces standard interface issues, preventing its inclusion in the comparative analysis. These limitations have been taken into account to ensure the consistency and validity of our experimental conditions.

\subsection{Result} \label{sec: result}

\subsubsection{Statistics Analysis}

Prior to presenting the results, it is crucial to reiterate the inherent challenges associated with solving POMDPs using model-free algorithms compared to model-based approaches. The primary difficulty arises from the lack of knowledge of the environment model in model-free algorithms. This includes the unknowns of the environment observation function, reward function, state transition function, and even the number of states and observations. The only available information for the model-free algorithm is the agent's action. Such limitations significantly amplify the complexity of solving POMDPs using model-free approaches.

Tab. \ref{tab: descriptive} presents a summary of descriptive statistics for the obtained results. It is observed that BIOMAP, QMDP, SARSOP, and POMDPSolver consistently achieve the globally optimal solution, regardless of changes in environmental parameters. Regarding the running time, while BIOMAP exhibits worse performance compared to QMDP and SARSOP, it is important to note that as a model-free algorithm, BIOMAP requires additional time for interacting with the environment and constructing the policy model. On the other hand, POMCPOW obtains the lowest cumulative reward among the algorithms, while POMDPSolver demonstrates the longest running time.


Moreover, we conducted an Analysis of Variance (ANOVA) \cite{st1989analysis} to determine if there is a significant difference between the two comparative groups presented in Tab. \ref{tab: anova}. The null hypothesis (H$0$) assumes that there is no significant relationship between group $1$ and group $2$, while the alternative hypothesis (H$1$) proposes that a significant relationship exists between the two groups \cite{fisher1966design}. By examining the P-value, we can determine if there is a significant difference. If the P-value indicates a significant difference, we reject the null hypothesis and accept the alternative hypothesis.

Tab. \ref{tab: anova} provides an analysis of the significance of the difference in means between the key variables of the experiment. ID $1$ confirms the significant differences in results observed among the various algorithms, reaffirming the findings from Tab. \ref{tab: descriptive}. IDs $2$-$5$ demonstrate that the mask parameter settings for the Masking cliff walking environment do not have an impact on the algorithm results. This indirectly showcases the reliability of the conclusions drawn for these algorithms. For ID $6$, the results indicate a substantial difference between BIOMAP and the other algorithms, with BIOMAP outperforming them. This finding establishes BIOMAP as a highly effective solution for solving the masking cliff walking experimental problem within the POMDP solver domain. Finally, ID $7$ provides evidence that the offline algorithm is more suitable for solving the masking cliff walking problem compared to the online algorithm. This suggests that finding the optimal strategy becomes challenging with a limited number of episodes when using the online algorithm.

These observations collectively highlight the strengths and advantages of BIOMAP while offering insights into the suitability of different algorithms for solving specific problems within the POMDP framework.

\subsubsection{Visualization}

\begin{figure}[t]
\centering
\subfloat[]{\includegraphics[width=0.445\linewidth]{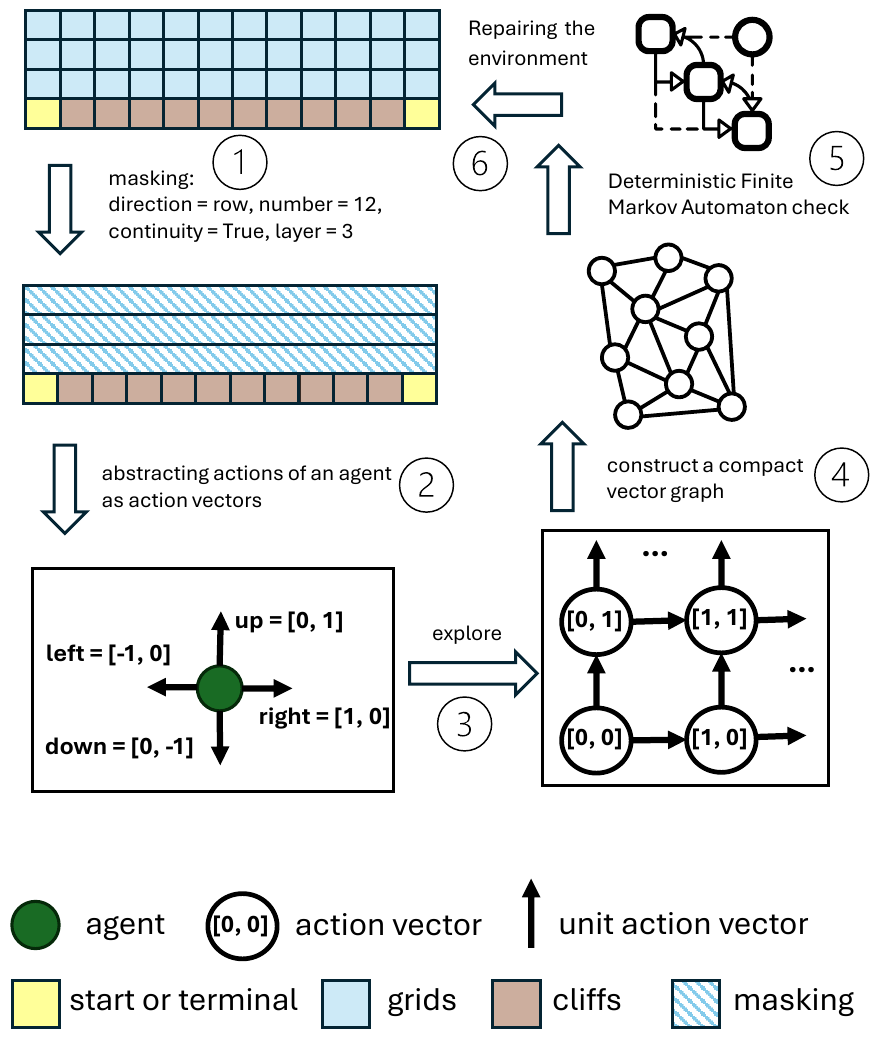}%
\label{fig: visualization1}}
\hfil
\subfloat[]{\includegraphics[width=0.445\linewidth]{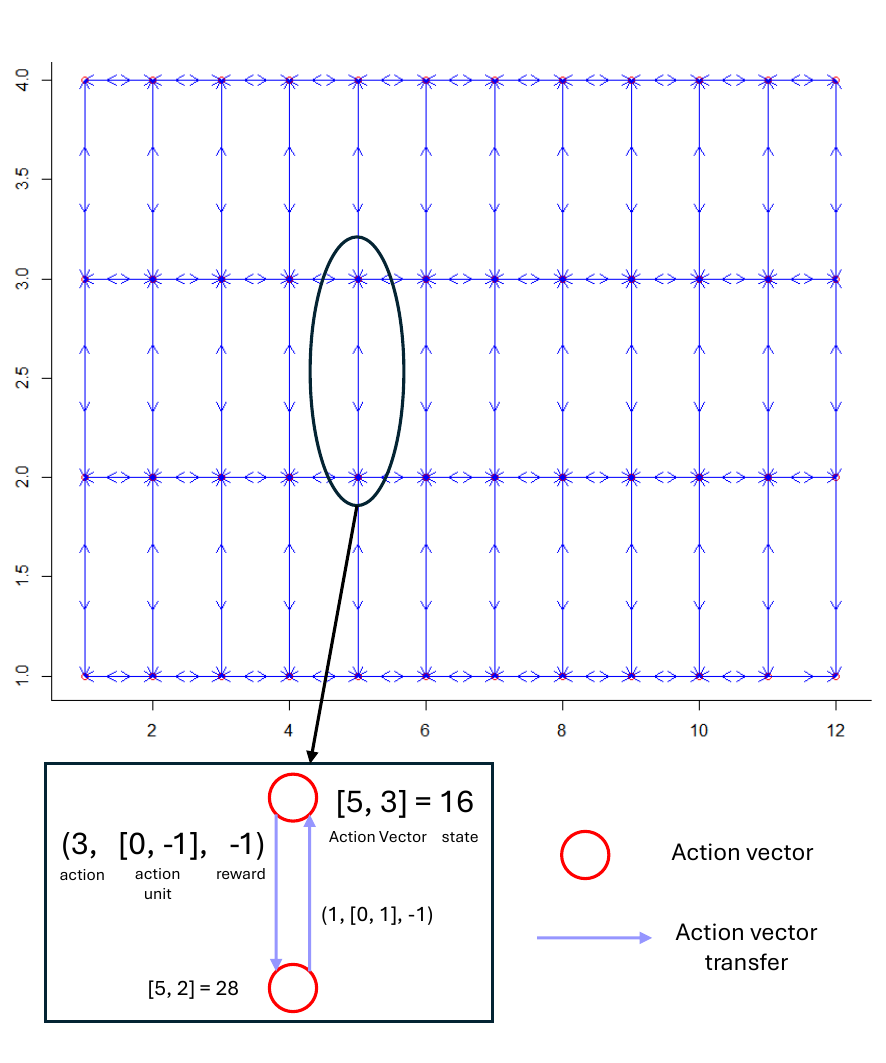}%
\label{fig: visualization2}}
\caption{Visualization of results. In (a), $\textcircled{1}$ is the process of experimental settings, $\textcircled{2} - \textcircled{6}$ are BIOMAP's working process on the masking cliff walking; (b) is a recovered Action Vector graph for Cliff Walking with masking (direction: row, number = $12$, continuity = True, layer = $3$) in (a).}
\label{fig: visualization}
\end{figure}

As mentioned previously, BIOMAP offers numerous advantages. Fig. \ref{fig: visualization} (a) presents the BIOMAP process for addressing DET-POMDP within a problematic environment. The process involves several steps: $\textcircled{1}$ masking the MDP: In the first step, the MDP is masked to incorporate the DET-POMDP scenario; $\textcircled{2}$ abstraction of agent's action: In the second step, the agent's action is abstracted and represented as a unit action vector; $\textcircled{3}$ exploration and graph recording: The third step involves exploring the environment with the agent, while simultaneously recording the action vector on the graph; $\textcircled{4}$ formation of Compact Vector Graph: As the exploration is completed, a Compact Vector Graph is formed, capturing the recorded actions and their relationships within the environment; $\textcircled{5}$ automaton detection: In the fifth step, the graph undergoes detection by an automaton, ensuring its adherence to predefined criteria; $\textcircled{6}$ reduction to MDP: Finally, in the sixth step, the graph is reconstructed to an MDP representation, allowing for subsequent analysis and solution generation. The BIOMAP process, as depicted in Fig. \ref{fig: visualization} (a), demonstrates the systematic approach for remediating a problematic environment affected by DET-POMDP.

Fig. \ref{fig: visualization} (b) provides a visual representation of the results obtained after the restoration process. The interpretation of both the vertex and edge components of the restored environment aligns with the definition outlined in Def. \ref{def:GMDP-graph}. Moreover, the overall structure of the repaired cliff walking environment is isomorphic and exhibits full observability.

\section{Conclusion}

In this paper, we addressed the DET-POMDP problem within the framework of POMDPs. We defined the challenge as a DET-POMDP, where the ambiguity in state-observation correspondence leads to inefficient policies, particularly in environments lacking explicit modeling. To effectively tackle this issue, we introduced the BIOMAP algorithm, a model-free biomimetics approach within an MDP-Graph-Automaton framework, transforming the DET-POMDP into a fully observable MDP. This enables the distinction between authentic and fraudulent environmental data, significantly enhancing the agent’s capacity to develop stable policies. Our results demonstrated BIOMAP's superior performance in maintaining operational effectiveness and environmental reparability compared to existing POMDP solvers, underlining its potential to improve the reliability of POMDP-based systems in fraud-prone environments. This research opens new avenues for deploying dependable autonomous systems in sectors susceptible to DET-POMDP, setting the stage for future advancements in reliable automated decision-making systems.

\bibliographystyle{unsrt}
\bibliography{main}

\begin{thebibliography}{10}

\bibitem{bonet2012deterministic}
Blai Bonet.
\newblock Deterministic pomdps revisited.
\newblock {\em arXiv preprint arXiv:1205.2659}, 2012.

\bibitem{hauskrecht1997planning}
Milos Hauskrecht.
\newblock {\em Planning and control in stochastic domains with imperfect information}.
\newblock PhD thesis, Massachusetts Institute of Technology, 1997.

\bibitem{kaelbling1998planning}
Leslie~Pack Kaelbling, Michael~L Littman, and Anthony~R Cassandra.
\newblock Planning and acting in partially observable stochastic domains.
\newblock {\em Artificial intelligence}, 101(1-2):99--134, 1998.

\bibitem{platt2010belief}
Robert Platt~Jr, Russ Tedrake, Leslie~Pack Kaelbling, and Tomas Lozano-Perez.
\newblock Belief space planning assuming maximum likelihood observations.
\newblock In {\em Robotics: Science and Systems}, volume~2, 2010.

\bibitem{cassandra1998survey}
Anthony~R Cassandra.
\newblock A survey of pomdp applications.
\newblock In {\em Working notes of AAAI 1998 fall symposium on planning with partially observable Markov decision processes}, volume 1724, 1998.

\bibitem{silver2010monte}
David Silver and Joel Veness.
\newblock Monte-carlo planning in large pomdps.
\newblock {\em Advances in neural information processing systems}, 23, 2010.

\bibitem{mnih2015human}
Volodymyr Mnih, Koray Kavukcuoglu, David Silver, Andrei~A Rusu, Joel Veness, Marc~G Bellemare, Alex Graves, Martin Riedmiller, Andreas~K Fidjeland, Georg Ostrovski, et~al.
\newblock Human-level control through deep reinforcement learning.
\newblock {\em nature}, 518(7540):529--533, 2015.

\bibitem{mnih2013playing}
Volodymyr Mnih, Koray Kavukcuoglu, David Silver, Alex Graves, Ioannis Antonoglou, Daan Wierstra, and Martin Riedmiller.
\newblock Playing atari with deep reinforcement learning.
\newblock {\em arXiv preprint arXiv:1312.5602}, 2013.

\bibitem{egorov2017pomdps}
Maxim Egorov, Zachary~N Sunberg, Edward Balaban, Tim~A Wheeler, Jayesh~K Gupta, and Mykel~J Kochenderfer.
\newblock Pomdps. jl: A framework for sequential decision making under uncertainty.
\newblock {\em Journal of Machine Learning Research}, 18(26):1--5, 2017.

\bibitem{10495179}
Eric Xia, Koulik Khamaru, Martin~J. Wainwright, and Michael~I. Jordan.
\newblock Instance-optimality in optimal value estimation: Adaptivity via variance-reduced q-learning.
\newblock {\em IEEE Transactions on Information Theory}, pages 1--1, 2024.

\bibitem{littman1995learning}
Michael~L Littman, Anthony~R Cassandra, and Leslie~Pack Kaelbling.
\newblock Learning policies for partially observable environments: Scaling up.
\newblock In {\em Machine Learning Proceedings 1995}, pages 362--370. Elsevier, 1995.

\bibitem{lovejoy1991computationally}
William~S Lovejoy.
\newblock Computationally feasible bounds for partially observed markov decision processes.
\newblock {\em Operations research}, 39(1):162--175, 1991.

\bibitem{pineau2003point}
Joelle Pineau, Geoff Gordon, Sebastian Thrun, et~al.
\newblock Point-based value iteration: An anytime algorithm for pomdps.
\newblock In {\em Ijcai}, volume~3, pages 1025--1032, 2003.

\bibitem{shani2013survey}
Guy Shani, Joelle Pineau, and Robert Kaplow.
\newblock A survey of point-based pomdp solvers.
\newblock {\em Autonomous Agents and Multi-Agent Systems}, 27:1--51, 2013.

\bibitem{cassandra2013incremental}
Anthony~R Cassandra, Michael~L Littman, and Nevin~Lianwen Zhang.
\newblock Incremental pruning: A simple, fast, exact method for partially observable markov decision processes.
\newblock {\em arXiv preprint arXiv:1302.1525}, 2013.

\bibitem{kocsis2006bandit}
Levente Kocsis and Csaba Szepesv{\'a}ri.
\newblock Bandit based monte-carlo planning.
\newblock In {\em European conference on machine learning}, pages 282--293. Springer, 2006.

\bibitem{chaslot2008monte}
Guillaume Chaslot, Sander Bakkes, Istvan Szita, and Pieter Spronck.
\newblock Monte-carlo tree search: A new framework for game ai.
\newblock In {\em Proceedings of the AAAI Conference on Artificial Intelligence and Interactive Digital Entertainment}, volume~4, pages 216--217, 2008.

\bibitem{sunberg2018online}
Zachary Sunberg and Mykel Kochenderfer.
\newblock Online algorithms for pomdps with continuous state, action, and observation spaces.
\newblock In {\em Proceedings of the International Conference on Automated Planning and Scheduling}, volume~28, pages 259--263, 2018.

\bibitem{bai2014integrated}
Haoyu Bai, David Hsu, and Wee~Sun Lee.
\newblock Integrated perception and planning in the continuous space: A pomdp approach.
\newblock {\em The International Journal of Robotics Research}, 33(9):1288--1302, 2014.

\bibitem{10399895}
Sagnik Bhattacharya and Prakash Narayan.
\newblock Shared information for a markov chain on a tree.
\newblock {\em IEEE Transactions on Information Theory}, pages 1--1, 2024.

\bibitem{kochenderfer2022algorithms}
Mykel~J Kochenderfer, Tim~A Wheeler, and Kyle~H Wray.
\newblock {\em Algorithms for decision making}.
\newblock MIT press, 2022.

\bibitem{kurniawati2009sarsop}
Hanna Kurniawati, David Hsu, and Wee~Sun Lee.
\newblock Sarsop: Efficient point-based pomdp planning by approximating optimally reachable belief spaces.
\newblock 2009.

\bibitem{somani2013despot}
Adhiraj Somani, Nan Ye, David Hsu, and Wee~Sun Lee.
\newblock Despot: Online pomdp planning with regularization.
\newblock {\em Advances in neural information processing systems}, 26, 2013.

\bibitem{wu2021adaptive}
Chenyang Wu, Guoyu Yang, Zongzhang Zhang, Yang Yu, Dong Li, Wulong Liu, and Jianye Hao.
\newblock Adaptive online packing-guided search for pomdps.
\newblock {\em Advances in Neural Information Processing Systems}, 34:28419--28430, 2021.

\bibitem{ross2007aems}
St{\'e}phane Ross, Brahim Chaib-Draa, et~al.
\newblock Aems: An anytime online search algorithm for approximate policy refinement in large pomdps.
\newblock In {\em IJCAI}, pages 2592--2598, 2007.

\bibitem{puterman2014markov}
Martin~L Puterman.
\newblock {\em Markov decision processes: discrete stochastic dynamic programming}.
\newblock John Wiley \& Sons, 2014.

\bibitem{yu2023measuring}
Yide Yu, Yan Ma, Yue Liu, Dennis Wong, Kin Lei, and Jos{\'e}~Vicente Egas-L{\'o}pez.
\newblock Measuring the state-observation-gap in pomdps: An exploration of observation confidence and weighting algorithms.
\newblock In {\em IFIP International Conference on Artificial Intelligence Applications and Innovations}, pages 137--148. Springer, 2023.

\bibitem{schmid1984individually}
Paul Schmid-Hempel.
\newblock Individually different foraging methods in the desert ant cataglyphis bicolor (hymenoptera, formicidae).
\newblock {\em Behavioral Ecology and Sociobiology}, 14:263--271, 1984.

\bibitem{muller1988path}
Martin M{\"u}ller and R{\"u}diger Wehner.
\newblock Path integration in desert ants, cataglyphis fortis.
\newblock {\em Proceedings of the National Academy of Sciences}, 85(14):5287--5290, 1988.

\bibitem{wittlinger2007desert}
Matthias Wittlinger, Ru\"udiger Wehner, and Harald Wolf.
\newblock The desert ant odometer: a stride integrator that accounts for stride length and walking speed.
\newblock {\em Journal of experimental Biology}, 210(2):198--207, 2007.

\bibitem{wehner2008desert}
R{\"U}DIGER WEHNER.
\newblock The desert ant's navigational toolkit: procedural rather than positional knowledge.
\newblock {\em Navigation}, 55(2):101--114, 2008.

\bibitem{wehner2003desert}
Rudiger Wehner.
\newblock Desert ant navigation: how miniature brains solve complex tasks.
\newblock {\em Journal of Comparative Physiology A}, 189:579--588, 2003.

\bibitem{balda1998animal}
Russell~P Balda, Irene~M Pepperberg, and Alan~C Kamil.
\newblock {\em Animal cognition in nature: the convergence of psychology and biology in laboratory and field}.
\newblock Academic Press, 1998.

\bibitem{jorge2008loft}
Paulo Jorge, In{\^e}s Silva, and Luis Vicente.
\newblock Loft features reveal the functioning of the young pigeon’s navigational system.
\newblock {\em Naturwissenschaften}, 95(3):223--231, 2008.

\bibitem{jorge2006strategies}
P~Jorge, L~Vicente, and W~Wiltschko.
\newblock What strategies do homing pigeons use during ontogeny?
\newblock {\em Behaviour}, 143(1):105--122, 2006.

\bibitem{stone2017anatomically}
Thomas Stone, Barbara Webb, Andrea Adden, Nicolai~Ben Weddig, Anna Honkanen, Rachel Templin, William Wcislo, Luca Scimeca, Eric Warrant, and Stanley Heinze.
\newblock An anatomically constrained model for path integration in the bee brain.
\newblock {\em Current Biology}, 27(20):3069--3085, 2017.

\bibitem{benhamou1997path}
Simon Benhamou.
\newblock Path integration by swimming rats.
\newblock {\em Animal Behaviour}, 54(2):321--327, 1997.

\bibitem{karp2010reducibility}
Richard~M Karp.
\newblock {\em Reducibility among combinatorial problems}.
\newblock Springer, 2010.

\bibitem{cormen2022introduction}
Thomas~H Cormen, Charles~E Leiserson, Ronald~L Rivest, and Clifford Stein.
\newblock {\em Introduction to algorithms}.
\newblock MIT press, 2022.

\bibitem{dijkstra2022note}
Edsger~W Dijkstra.
\newblock A note on two problems in connexion with graphs.
\newblock In {\em Edsger Wybe Dijkstra: His Life, Work, and Legacy}, pages 287--290. 2022.

\bibitem{sutton2018reinforcement}
Richard~S Sutton and Andrew~G Barto.
\newblock {\em Reinforcement learning: An introduction}.
\newblock MIT press, 2018.

\bibitem{thrun1994learning}
Sebastian Thrun.
\newblock Learning to play the game of chess.
\newblock {\em Advances in neural information processing systems}, 7, 1994.

\bibitem{wiering2000multi}
Marco~A Wiering et~al.
\newblock Multi-agent reinforcement learning for traffic light control.
\newblock In {\em Machine Learning: Proceedings of the Seventeenth International Conference (ICML'2000)}, pages 1151--1158, 2000.

\bibitem{bechar2016agricultural}
Avital Bechar and Cl{\'e}ment Vigneault.
\newblock Agricultural robots for field operations: Concepts and components.
\newblock {\em Biosystems Engineering}, 149:94--111, 2016.

\bibitem{bezanson2017julia}
Jeff Bezanson, Alan Edelman, Stefan Karpinski, and Viral~B Shah.
\newblock Julia: A fresh approach to numerical computing.
\newblock {\em SIAM review}, 59(1):65--98, 2017.

\bibitem{team2013r}
R~Core Team.
\newblock R: A language and environment for statistical computing. r foundation for statistical computing.
\newblock {\em (No Title)}, 2013.

\bibitem{sondik1971optimal}
Edward~Jay Sondik.
\newblock {\em The optimal control of partially observable Markov processes}.
\newblock Stanford University, 1971.

\bibitem{cheng1988scaling}
Z~Cheng and S~Redner.
\newblock Scaling theory of fragmentation.
\newblock {\em Physical review letters}, 60(24):2450, 1988.

\bibitem{littman1994witness}
Michael~L Littman.
\newblock The witness algorithm: Solving partially observable markov decision processes.
\newblock {\em Brown University, Providence, RI}, 1994.

\bibitem{cassandra1997incremental}
Anthony Cassandra, Michael~L. Littman, and Nevin~L. Zhang.
\newblock Incremental pruning: a simple, fast, exact method for partially observable markov decision processes.
\newblock In {\em Proceedings of the Thirteenth Conference on Uncertainty in Artificial Intelligence}, UAI'97, page 54–61, San Francisco, CA, USA, 1997. Morgan Kaufmann Publishers Inc.

\bibitem{cassandra2004tony}
Anthony~R. Cassandra.
\newblock Tony's pomdp file repository page, 2004.

\bibitem{st1989analysis}
Lars St, Svante Wold, et~al.
\newblock Analysis of variance (anova).
\newblock {\em Chemometrics and intelligent laboratory systems}, 6(4):259--272, 1989.

\bibitem{fisher1966design}
Ronald~Aylmer Fisher, Ronald~Aylmer Fisher, Statistiker Genetiker, Ronald~Aylmer Fisher, Statistician Genetician, Great Britain, Ronald~Aylmer Fisher, and Statisticien G{\'e}n{\'e}ticien.
\newblock {\em The design of experiments}, volume~21.
\newblock Oliver and Boyd Edinburgh, 1966.

\bibitem{kameda1970state}
Tsunehiko Kameda and Peter Weiner.
\newblock On the state minimization of nondeterministic finite automata.
\newblock {\em IEEE Transactions on Computers}, 100(7):617--627, 1970.

\bibitem{hopcroft2001introduction}
John~E Hopcroft, Rajeev Motwani, and Jeffrey~D Ullman.
\newblock Introduction to automata theory, languages, and computation.
\newblock {\em Acm Sigact News}, 32(1):60--65, 2001.

\end{thebibliography}

\newpage

\appendix

\section{Summary Table for Sec. \ref{sec: related works}}


\begin{table}[h]
    \centering
    \caption{POMDP Solver Classification.}
    \begin{tabular}{|c|c|c|c|}
    \hline
    \tiny Algorithm & \tiny Classification & \tiny Advantage & \tiny Disadvantage \\
    \hline
     \tiny QMDP \cite{littman1995learning} ($1995$) & \tiny \multirow{4}{*}{\makecell{Value episode-based \\ Algorithms}} & \tiny \multirow{4}{*}{\makecell{Explicit Convergence Criteria,\\ Solid theoretical Foundations,\\ Guaranteed Policy Improvement}} & \tiny \multirow{4}{*}{\makecell{High-Dimensional Challenge,\\ High Computational Cost}}\\
    \tiny Belief Grid Value episode \cite{lovejoy1991computationally} ($1991$) & & & \tiny \multirow{3}{*}{} \\
    \tiny PBVI \cite{pineau2003point} \cite{shani2013survey} ($2003$)  & & & \\
    \tiny Incremental Pruning \cite{cassandra2013incremental} ($2013$) & & & \\ 
    \hline
    
    \tiny PO-UCT \cite{kocsis2006bandit} ($2010$) & \tiny \multirow{4}{*}{Monte Carlo Algorithms} & \tiny \multirow{4}{*}{\makecell{Applicability to High-Dimensional Spaces,\\ Adaptability}} & \tiny \multirow{4}{*}{\makecell{Sample Dependence,\\ Large Computational Requirement,\\ Variance Issues}} \\
    \tiny POMCPOW \cite{sunberg2018online} ($2018$) & & & \\
    \tiny MCVI \cite{bai2014integrated} ($2014$) & & & \\
    \tiny PFT-DPW \cite{sunberg2018online} (-) & & &\\
    \hline
    
     \tiny FIB \cite{kochenderfer2022algorithms} ($2022$) & \tiny \multirow{4}{*}{Tree Search-based Algorithms} & \tiny \multirow{4}{*}{\makecell{Effective Online Decision-Making,\\ Strong Adaptability,\\ Structured Exploration}} & \tiny \multirow{4}{*}{\makecell{Branch Management,\\ Large Computational Requirement,\\ Local Optima Traps}} \\
     \tiny SARSOP \cite{kurniawati2009sarsop} ($2008$) & & & \\
     \tiny AR-DESPOT \cite{somani2013despot} ($2013$) &  & & \\
     \tiny AdaOPS \cite{wu2021adaptive} ($2021$) & & & \\
     \hline
     
     \tiny AEMS \cite{ross2007aems} ($2007$) & \tiny \makecell{Heuristic and Approximation \\ Algorithm} & \tiny \makecell{Rapid Solution Generation,\\ Heuristic Guidance,\\ Real-Time Application Capability} & \tiny \makecell{No Guarantee of Global Optimum,\\ Heuristic Quality Dependence,\\ Empirical Nature} \\
    \hline
    \end{tabular}
    \label{tab: related works}
\end{table}

\section{Bionics Mapping Relation Table for Sec. \ref{sec: bionic}}

In conclusion, we summarize the Bionics mapping relation in Tab. \ref{tab: bionic}.

\begin{table}[h!]
    \centering
    \caption{Bionics mapping relation.}
    \begin{tabular}{|c|c|}
    \hline
    \small \textbf{Desert Ant Ability} & \small \textbf{BIOMAP} \\
    \hline
     \small Gait and Step Frequency & \small Action Unit Vector \\
     \small Internal Pedometer & \small Action Vector\\
     \small Angle Monitoring & \small Action Vector\\
     \small Memory and Learning & \small Compact Vector Graph \\
     \small Path Integration &  \small Euclidean norm\\
     \small Celestial Compass & \small Boundary Arbiter\\
     \hline
    \end{tabular}
    \label{tab: bionic}
\end{table}

\section{Background} \label{app-background}

\subsection{MDPs and POMDPs} \label{sec: mdps and pomdps}

\begin{definition}[Markov Decision Process] \label{def: mdp}
A Markov Decision Process can be consist with 5-tuple $\mathbb{M} =\langle \mathcal{S}, \mathcal{A}, \mathcal{T}, \mathcal{R}, \gamma \rangle$. $\mathcal{S}$ is a finite state space, and $\mathcal{A}$ denotes a finite action space. For non-deterministic MDP, $\mathcal{T}(s, a) = s^\prime$ represents the state transition function that the state $s$ transfers to the other state $s^\prime$ after taking action $a$, and the transition probability $1 \geq P(s^\prime \mid s, a) > 0$. $\mathcal{R}(s, a, s^\prime)$ is the immediate reward when the state $s$ moves to $s^\prime$ after taking action $a$. For MDP with a deterministic environment, the state transition probability is $P(s^\prime \mid s, a) = 1$, and the reward function is $\mathcal{R}(s, a)$. $\gamma \in [0, 1]$ is a discount factor.
\end{definition}

The whole process of MDP manifests the recurrence property along with time. An agent initially starts with a state $s_t \in \mathcal{S}$ at time-step $t \in \mathcal{T}$, after taking action $a_t \in \mathcal{A}$, then the agent receives a reward $r_t \in \mathcal{R}$ and moves to the next state $\mathcal{T}(s, a) = s^\prime$. If this is a deterministic MDP, the state transition probability is fixed $P(s^\prime \mid s, a) = 1$. Otherwise, the probability is unstable, and the range of it is $1 \geq P(s^\prime \mid s, a) > 0$. Iteratively, the state $s_{t+1}$ becomes the current state for the agent, and the procedure mentioned above is executed circularly along with time series $t \in T$.

The value of states denotes by $V^\pi(s)$ that:

\begin{equation}
    V^\pi(s) \triangleq \mathbb{E}_\pi [\sum^\infty_{t=0} \gamma^t R(s_t, \pi(s_t)) \mid s_0 = s].
\end{equation}

Also, the value of state-action pairs can be measured by Q-value $Q(s, a)$. According to the Bellman Equation, the Q-value is defined as follows:

\begin{equation}
    Q_{t+1}(s, a) = R(s, a) + \gamma \sum_{s^\prime \in \mathcal{S}} [P(s, a, s^\prime) Q_{t}(s^\prime, a^\prime)].
\end{equation}

POMDPs \cite{kaelbling1998planning} are an extension of MDPs. In POMDPs, the agent does not have complete visibility of the states in the environment. Instead, it can only observe partial states. These uncertain observations introduce ambiguity in determining the optimal decision.

\begin{definition}[POMDPs]
    In general, a POMDP consists of a tuple $\mathbb{P} = \langle \mathcal{S}, \mathcal{A}, \mathcal{O}, \Omega, R, \mathcal{T}, \gamma \rangle$. At each time step $t \in \mathcal{T}$, the agent takes an action $a_t \in \mathcal{A}$ at the state $s_t \in \mathbb{S}$. Then the environment transitions to a new state $s_{t+1}$ by the state transition function $\mathcal{T}: \mathcal{S} \times \mathcal{A} \times \mathcal{S} \rightarrow [0,1]$. After that, the environmental feedback is the observation $o_t \in \mathcal{O}$ by the observation function $\Omega: \mathcal{S} \rightarrow \mathcal{O}$ and the reward $r_t$ by the reward function $R: \mathcal{S} \times \mathcal{A} \rightarrow \mathcal{R}$. $\gamma \in [0,1]$ is a discount factor.
\label{def:pomdp}
\end{definition}

\subsection{Automata}

Automata and formal language provide a mathematical framework for understanding and analyzing the behaviors of computer systems. They also enable the creation of languages that can be recognized by computers.

Automata are mathematical models of computers that consist of sequential states and transitions. They can exhibit either deterministic or non-deterministic behavior. There are three common categories of automata: Finite Automata (FA), Pushdown Automaton (PDA), and Turing Machine (TM). This paper specifically focuses on Finite Automata (FA).

The definitions of Deterministic Finite Automaton (DFA) and Non-deterministic Finite Automaton (NDFA), as described by Kameda et al. \cite{kameda1970state}, are presented below:

\begin{definition}[DFA \& NDFA]
    A (deterministic) automaton is a system $M_{Markov}=\langle Q, \Sigma, \delta, q_0, F \rangle$, where
    \begin{itemize}
        \item $Q$ is a finite non-empty set, the set of states,
        \item $\Sigma$ is a finite alphabet, the input alphabet,
        \item $q_0 \in Q$ is a designated state, the initial state,
        \item $F \subseteq Q$ is the set of final states,
        \item $\delta: Q \times \Sigma \rightarrow Q$ is mapping, the transition function.
    \end{itemize}
    If $Q_0=\left\{q_0\right\}$, and if $|\delta(q, \sigma)|=1$ for $\forall q \in Q$ and $\forall \sigma \in \Sigma$, then the automaton $\delta$ is said to be a \textit{deterministic finite automaton} (DFA). Otherwise, $\delta$ is said to be a \textit{nondeterministic automation} (NDFA), which denotes $M^\prime$.
\label{def:automaton}
\end{definition}

Moreover, language is an important part of Automata. First, we cite the definition of automaton configurations.

\begin{definition}[Configurations of Automaton \cite{hopcroft2001introduction}]
    The set $C=Q \times \Sigma^*$ is the set of configurations of $M$. A configuration $\left(q, a_1 \ldots a_n\right)$ is interpreted in the way that $M$ is in state $q$ and gets the word $a_1 \ldots a_n$ as input.
\label{def:configurations}
\end{definition}

The transition relation, denoted by $\vdash_M$, is a binary relation over $C$ and is defined in the following way.

\begin{definition}[Transition Relation \cite{hopcroft2001introduction}]
    For every $(q, w),\left(q^{\prime}, w^{\prime}\right) \in C$, we have $(q, w) \vdash_M\left(q^{\prime}, w^{\prime}\right)$ if and only if $w=a w^{\prime}$, for some $a \in \Sigma$ and the inclusion $q^{\prime}=\delta(q, a)$ holds.
\label{def: transition relation}
\end{definition}

After that, we have the definition of the language recognized by $M$.

\begin{definition}
    Let $M=\left(Q, \Sigma, \delta, q_0, F\right)$ be an automaton. The language recognized by $M$ is 
    $$
    L(M)=\left\{w \in \Sigma^* \mid\left(q_0, w\right) \vdash_M^*(q, \varepsilon) \text { for some } q \in F\right\}
    $$
\label{def:recognized}
\end{definition}

\newpage

\section{Proof of Thm. \ref{theorem: variance} \label{proof: theorem: variance}}
\begin{proof}
    In a DET-POMDP problem $\mathcal{F}$, $\exists s_1, s_2, \cdots, s_n \in \mathcal{S}$ share the same state information. Set $Q(s_1, a), Q(s_2, a), \cdots Q(s_n, a)$ are the real Q-value of them for a same action $\forall a \in \mathcal{A}$. According to Def. \ref{def: env}, the environment deceives the agent that all these states share the same state information. Thus we have the estimated Q-value that $\hat{Q}(s_1, a) = \hat{Q}(s_2, a) = \cdots = \hat{Q}(s_n, a)$, and we use $\hat{Q}(s, a)$ to represent them.

    Let the error term $e_i$ denote the difference between $i$-th real and estimated Q-value:
    \begin{equation}
        e_i=\hat{Q}(s, a)-Q^*\left(s_i, a\right).
    \label{equ: error-term}
    \end{equation} The variance of the error term is
    \begin{equation}
    \operatorname{Var}(e)=\frac{1}{n} \sum_{i=1}^n\left(e_i-\bar{e}\right)^2,
    \label{equ: error-term-variance}
    \end{equation} where the $\bar{e}$ is the average of error terms, which represents as
    \begin{equation}
        \bar{e}=\frac{1}{n} \sum_{i=1}^n e_i.
    \label{equ: average-q-value}
    \end{equation}
    
    Because all states are observed in the same observation, we combine Equ. \ref{equ: error-term} and \ref{equ: average-q-value} as
    \begin{equation}
        \bar{e}=\hat{Q}(s, a)-\frac{1}{n} \sum_{i=1}^n Q\left(s_i, a\right).
    \end{equation} Therefore, we re-write the variance Equ. \ref{equ: error-term-variance} as
    \begin{equation}
        \operatorname{Var}(e) = \frac{1}{n} \sum^n_{i=1} (\hat{Q}(s, a) - Q(s_i, a) - \bar(e))^2 .
    \end{equation}
    The extension of the variance equation is
    \begin{equation}
        \begin{aligned}
        \operatorname{Var}(e) = \frac{1}{n} \sum^n_{i=1} (\hat{Q}(s, a) &- Q(s_i, a) - \hat{Q}(s, a) \\
        &+ \frac{1}{n} \sum^n_{j=1} Q(s_j, a))^2
    \end{aligned}
    \end{equation}

    \begin{equation}
        \operatorname{Var}(e) = \frac{1}{n} \sum^n_{i=1} (-Q(s_i, a) + \frac{1}{n} \sum^n_{j=1} Q(s_j, a))^2
    \end{equation}

\end{proof}

\newpage
	
\section{Proof of Lem. \ref{lemma:mdp to dfma} \label{proof: lemma:mdp to dfma}} 

\begin{proof}
Let $\mathcal{H}^a = \{h_1^a, h_2^a, \cdots, h_n^a\}$, where $n > 0$ (from Def. \ref{def: mdp}). $\exists h^a$, $h^a = (a_t, a_{t+1}, \cdots, a_{t+k} \in \mathcal{H}^a$ and $\circ h^a = \{a_t, a_{t+1}, \cdots, a_{t+m}) \subseteq \Sigma$, where $k \geq m > 0$. $\forall a \in A$ and $\forall s \in S$ (from Def. \ref{def: mdp}). $q_i = s_i$.\\

Then, $L_{Markov} = \{h_1^a, h_2^a, \cdots, h_n^a\}$ because $\mathcal{H}^a = L_{Markov}$ (according to Def. \ref{def:automaton}).
Because $\forall a \in \exists \circ h^a$. And because $A \subset \Sigma$ (from Def. \ref{def:automaton}). So, we get the 
\begin{equation}
    \forall a \in \exists \circ h^a \subset \Sigma.
\label{equ: s,a in Sigma}
\end{equation}

For $\forall a \in \Sigma$, let $w = (a_t, a_{t+1}, \cdots, a_{t+k})$ be a word. Set $w \in L_{Markov}$, and we know $\forall h \in \mathcal{H}^a$. Because $\mathcal{H}^a = L_{Markov}$, then we have

\begin{equation}
    \exists w = \exists h
\label{equ:w in h}
\end{equation}

If $t=1$, for $a_1 \in \Sigma$, let $w_0 = (a_0, a_1)$ and $w_1 = (a_1)$, and $(s_0, w_0), (s_1, w_1) \in C$. Then we have $w_1 = a_0 w_0$. And we get $w_1 = \delta(s_0, a_0)$. Thus, we get $(s_0, w_0) \vdash \left(s_1, w_1\right)$ (from Def. \ref{def: transition relation}).\\

If $t = k$ and $k>1$, because $\mathcal{T}: S \times A \rightarrow S$ (according to Def. \ref{def: mdp}) and $S \subset Q$, $A \subset \Sigma$ (according to Def. \ref{def:automaton}), thus we have the derivation chain  $\delta (s_t,(a_t, a_{t+1}, \cdots, a_{t+k})) \vdash \delta (s_{t+1},(a_{t+1})), \cdots, a_{t+k})) \vdash \cdots \vdash \delta (s_{t+k},(a_{t+k}))$. Then we have $(q_0, w) \vdash (q, \varepsilon)$, where $q \in F$. Finally, we have the $L_{Markove}$ recognized by $M_{Markov}$ (from Def. \ref{def:recognized}).

\end{proof}

\newpage

\section{Proof of Lem. \ref{lemma:pomdp to ndfma} \label{proof: lemma:pomdp to ndfma}} 

\begin{proof}

According to Lem. \ref{lemma:mdp to dfma}, we know the action trajectory of a MDP recognizable $M$ by a Deterministic Finite Markov Automaton $M_{Markov}$. \\

Let $A = A^\prime$ and $S = S^\prime$. $\mathcal{H}^{a\prime} = \{h_1^a, h_2^a, \cdots, h_n^a\}$, where $n > 0$ (from Def. \ref{def: mdp}). $Q^\prime = S^\prime$\\

In a Fully MDP $M$, because the observable state is real, $P(s \mid o) = 1$, where $s \in S$ and $o \in \mathcal{O}$. (from Def. \ref{def: mdp}) In a POMDP, $P(s \mid o) \leq 1$ is the probability of the state $s$ arrived if get the observation $o$ (from Def. \ref{def:pomdp} and Measuring the State-Observation-Gap in POMDPs: An Exploration of Observation Confidence and Weighting Algorithms). Thus, $P(s, a \mid o) \leq 1$, then we have 
\begin{equation}
    \mid \delta (q, \sigma) \mid \leq 1
\label{equ: ndfma transition smaller than 1}
\end{equation}
Because Equ. \ref{equ: ndfma transition smaller than 1} satisfies the definition of NDFMA (Def. \ref{def:automaton}), we have the language $L_{Markov}^\prime = H^\prime$ recognizable by the Nondeterministic Finite Markov Automaton $M_{Markov}^\prime$.

\end{proof}

\end{document}